\documentclass[showpacs,prl,onecolumn,aps,superscriptaddress,preprintnumbers,nofootinbib]{revtex4}
\usepackage[T1]{fontenc}
\usepackage[latin9]{inputenc}
\usepackage{amsmath,amssymb,bigints}
\usepackage{epsfig}
\usepackage{graphicx}
\usepackage{amsmath}
\usepackage{amsfonts}
\usepackage{changes}

\renewcommand{\vec}{\boldsymbol}
\newcommand{\beq}{\begin{equation}}
\newcommand{\eeq}{\end{equation}}
\newcommand{\bea}{\begin{eqnarray}}
\newcommand{\eea}{\end{eqnarray}}
\newcommand{\baa}{\begin{array}}
\newcommand{\eaa}{\end{array}}
\renewcommand{\vec}[1]{\boldsymbol{#1}}
\def\eq#1{{Eq.~(\ref{#1})}}
\def\fig#1{{Fig.~\ref{#1}}}

\newcommand{\bas}{\bar{\alpha}_S}
\newcommand{\as}{\alpha_S}

\newcommand{\nn}{\nonumber}

\newcommand{\h}{\frac{1}{2}}

\newcommand{\ga}{\gamma}

\newcommand{\De}{\Delta}

\newcommand{\Lb}{\left(}
\newcommand{\Rb}{\right)}
\def\pom{{I\!\!P}}

\renewcommand{\vec}[1]{\boldsymbol{#1}}

\def\pom{{I\!\!P}}

\newcommand{\kmin}{\kappa_{\min}}
\newcommand{\kmax}{\kappa_{\max}}
\newcommand{\mpi}{\tilde{m}}      
\newcommand{\mmi}{{\tilde{m}^*}}  
\newcommand{\mzi}{{\tilde{m}_0}}  

\begin{document}

\title{High energy evolution for Gribov-Zwanziger confinement: solution to the equation}
\author{E. ~Gotsman}
  \email{gotsman@post.tau.ac.il}
  \affiliation{Department of Particle Physics, School of Physics and Astronomy,
    Raymond and Beverly Sackler Faculty of Exact Science,
    Tel Aviv University, Tel Aviv, 69978, Israel}
\author{Yu.~P.~ Ivanov}
\email{yuri.ivanov@usm.cl}
  \affiliation{Joint Institute for Nuclear Research (JINR), Dubna, 141980 Moscow Region, Russia}
  \affiliation{Departamento de F\'isica, Universidad T\'ecnica Federico Santa Mar\'ia
    and Centro Cient\'ifico Tecnol\'ogico de Valpara\'iso,
    Casilla 110-V, Valparaiso, Chile}
\author{E.~ Levin}
  \email{leving@tauex.tau.ac.il, eugeny.levin@usm.cl}
  \affiliation{Department of Particle Physics, School of Physics and Astronomy,
    Raymond and Beverly Sackler Faculty of Exact Science,
    Tel Aviv University, Tel Aviv, 69978, Israel}
  \affiliation{Departamento de F\'isica, Universidad T\'ecnica Federico Santa Mar\'ia
    and Centro Cient\'ifico Tecnol\'ogico de Valpara\'iso,
    Casilla 110-V, Valparaiso, Chile}

\date{\today}


\begin{abstract}
In this paper we solved the new evolution equation for high energy scattering amplitude
that stems from the Gribov-Zwanziger approach to the confinement of quarks and gluons.
We found that (1) the energy dependence of the scattering amplitude turns out to be
the same as for QCD BFKL evolution; (2) the spectrum of the new equation does not depend
on the details of the Gribov-Zwanzinger approach and (3) all eigenfunctions coincide with
the eigenfunctions of the QCD BFKL equation at large transverse momenta $\kappa\,\geq\,1$.
The numerical calculations show that there exist no new eigenvalues with the eigenfunctions
which decrease faster than solutions of the QCD BFKL equation at large transverse momenta.
The structure of the gluon propagator in Gribov-Zwanziger approach, that stems from the
lattice QCD and from the theoretical evaluation, results in the exponential suppression
of the eigenfunctions at long distances and in the resolution of the difficulties, which
the Colour Glass Condensate (CGC) and some other approaches, based on perturbative QCD,
face at large impact parameters. We can conclude that the confinement of quark and gluons,
at least in the form of Gribov-Zwanziger approach, does not influence on the scattering
amplitude except solving the long standing theoretical problem of its behaviour at large
impact parameters.

\end{abstract}

\preprint{}
\maketitle
\vspace{-1cm}
\tableofcontents
\newpage

\section{ Introduction}

It is well known that perturbative QCD suffers a fundamental problem: the scattering
amplitude decreases at large impact parameters ($b$) as a power of $b$. Such behaviour
contradicts the Froissart theorem\cite{FROI} and, hence, perturbative QCD cannot lead
to an effective theory at high energy.

In particular, the CGC/saturation approach (see Ref.\cite{KOLEB} for a review) which
is based on perturbative QCD, is confronted by this problem \cite{KW,FIIM}. At large
$b$ the scattering amplitude is small and, therefore, only the linear BFKL (Balitsky,
Fadin, Kuraev and Lipatov) equation\cite{BFKL} describes the scattering amplitude in
perturbative QCD. It is known that the eigenfunction of this equation (the scattering
amplitude of two dipoles with sizes $r$ and $R$) has the following form\cite{LIP}
\beq\label{EIGENF}
\phi_\gamma\Lb \vec{r}, \vec{R}, \vec{b}\Rb\,\,=\,\,
  \Lb \frac{r^2\,R^2}{\Lb \vec{b} + \h(\vec{r} - \vec{R})\Rb^2\,\Lb \vec{b} - \h(\vec{r} - \vec{R})\Rb^2}
  \Rb^\gamma\,\,\xrightarrow{b\,\gg\,r,R}\,\,\Lb \frac{r^2\,R^2}{b^4}\Rb^\gamma
\eeq

One can see that $\phi_\gamma\Lb \vec{r}, \vec{R}, \vec{b}\Rb$ at large impact parameter
$b$ decreases as a power of $b$. In particular, such a decrease leads to the growth of
the radius of interaction as a power of the energy\cite{KW,FIIM}, resulting in the violation
of Froissart theorem. Since it was proven in Ref.\cite{LIP} that the eigenfunction of any
kernel with conformal symmetry has the form of \eq{EIGENF}, we can only change the large
$b$ behaviour by introducing a new dimensional scale in the kernel of the equation.
A variety of ideas to overcome this problem have been suggested in Refs.\cite{LERYB1,LERYB2,LETAN,QCD2,KHLEP,KKL,FIIM,GBS1,BLT,GKLMN,HAMU,MUMU,BEST1,BEST2,KOLE,LETA,LLS,LEPION,KHLE,KAN,GOLEB}.
In our previous paper \cite{GOLEM} we used the Gribov-Zwanziger approach\cite{GRI0,GRI1,GRI2,GRI3,GRI4,PVB,Z1,Z2,Z3,GRREV,DOKH}
for the confinement of quarks and gluons to fix this non-perturbative scale. We derived
the generalized BFKL evolution equation, which incorporates this new dimensional scale,
and demonstrated that this equation leads to the exponential decrease of the scattering
amplitude at large $b$. We will discuss both the equation and large $b$ behaviour of
the solution in the next section which has a review character.

The goal of this paper is to find the solution to this new equation. In section III
we consider the general properties of the spectrum and eigenfunctions, which follow
from an analytical approach. In particular, we prove that the eigenfunctions of \eq{EIGENF}
describe the eigenfunctions of the new equation at short distances $r\,\,\ll\,\,R$.
In section IV we concentrate our efforts on the numerical solution of the equation.
We show that all eigenvalues of the new equation, that generate the power energy
increase of the scattering amplitude, coincide with the massless BFKL eigenvalues.
However the eigenfunctions have quite a different behaviour in comparison with the
eigenfunction of the massless BFKL equation and they crucially depend on the input
from Gribov-Zwanziger confinement approach. Finally, in Section V we discuss our
results and future prospects.

\section{BFKL evolution equation for Gribov-Zwanziger confinement - a recap}

\subsection{Gribov - Zwanziger confinement: gluon propagator}

As we have alluded that in Ref.\cite{LIP} it is proven, that eigenfunctions of \eq{EIGENF}
have the same form for all kernels with conformal symmetries. Hence we have to modify the
kernel of the BFKL equation introducing a new dimensional scale of the non-perturbative
origin. In other words, we need an approach which models the confinement of quarks and
gluons. Among numerous approaches to confinement, the one proposed by Gribov,
\cite{GRI0,GRI1,GRI2,GRI3,GRI4,PVB,Z1,Z2,Z3,GRREV,DOKH} has special advantages,
which makes it most suitable for discussion of the BFKL equation in the framework
of this hypotheses. First, it is based on the existence of Gribov copies \cite{GRI0}
- multiple solutions of the gauge-fixing conditions, which are the principle properties
of non-perturbative QCD. Second, the main ingredient is the modified gluon propagator,
which can be easily included in the BFKL-type of equations. Third, in Ref.\cite{KHLE}
(see also ref.\cite{FDGS}) it is demonstrated that the Gribov gluon propagator originates
naturally from the topological structure of non-perturbative QCD in the form:
\beq \label{GLPR}
  G\Lb q\Rb\,\,=\,\,\frac{1}{q^2\,+\,\,\frac{\chi_{\rm top}}{q^2}}\,\,=\,\,\frac{q^2}{q^4\,+\,\mu^4}
\eeq
where $\chi_{\rm top}\,=\,\mu^4 $ is the topological susceptibility of QCD, which
is related to the $\eta'$ mass by the Witten-Veneziano relation\cite{VEN,WIT}. This
allows us to obtain the principal non-perturbative dimensional scale, directly from
the experimental data.

However, it is shown in Ref.\cite{GOLEM} that propagator of \eq{GLPR}, which vanishes at
$q=0$, does not lead to the exponential suppression of the scattering amplitude at large
impact parameters ($b$). Fortunately, the lattice calculation of the gluon propagator
generates the gluon propagator with $G\Lb q=0\Rb\,\neq\,0$ (see Refs.\cite{DOS,DOV,CDMV}
and references therein), in explicit contradiction with \eq{GLPR}.

In Refs.\cite{HU1,CFPS,CFMPS,CDMV,DSV,HU2,HU3,AHS,DOV,GRA,FMP,DSVV,DGSVV,CLSST,Z4,Z5,LVS}%
\footnote{This list of references is not complete. More details can be found  in 
the reviews \cite{GRREV,HU1}.} it is shown that $G\Lb q=0\Rb\,\neq\,0$ is a general
feature of non-perturbative approaches and that Gribov's copies lead to the gluon
propagator which is final at $q \to 0$. In this paper we parameterize the gluon
propagator in the following form:
\beq \label{GGLPR1}
  G\Lb q\Rb \,\,=\,\,\frac{q^2\,+\,M^2_0}{\Lb q^2\,+\,M^2\Rb^2\,+\,\mu^4}
\eeq

We view this form as parameterization of the sum of Gribov's propagators of \eq{GLPR}
with different values of $\mu$, as it has been discussed in Ref.\cite{GOLEM}. We are
aware that \eq{GGLPR1}, which describes the lattice QCD data, is a simplified version
of the refined Gribov-Zwanziger (RGZ) theoretical approaches that have been discussed
in Refs.\cite{HU1,CFPS,CFMPS,CDMV,DSV,HU2,HU3,AHS,DOV,GRA,FMP,DSVV,DGSVV,CLSST,Z4,Z5,LVS}.
However, we believe that it is a good first approximation, which allows us to introduce
two dimensional parameters from confinement physics. In this paper we call the gluon
propagator of \eq{GGLPR1} as the lattice QCD propagator or as RGZ propagator.

As we have mentioned, at high energies $q$ is a two dimensional vector, which corresponds
to transverse momentum carried by the gluon. Introducing
\beq \label{GGLPR2}
  G^{\pm}\Lb q\Rb\,\,=\,\,\frac{1}{\Lb q^2\,+\,M^2\Rb\,\pm\,i\,\mu^2}
\eeq
we can re-write \eq{GGLPR1} in the form:
\bea \label{GGLPR3}
  G(q)&=&\h\Big(G^{+}(q)+G^{-}(q)\Big)\,+\,\frac{M^2_0-M^2}{2\,i\mu^2}
           \Big(G^{+}(q)-G^{-}(q)\Big)
     \,=\,\frac{1}{\mu^2}\Big({\rm Re}\,G^{+}(\kappa)\,+\,(M^2_0-M^2)\,{\rm Im}\,G^{+}(\kappa)\Big)\\
      &=&\h\Bigg\{\mkern-8mu
           \Lb 1 + i\,\frac{M^2_0\!-\!M^2}{\mu^2}\Rb G^{+}(q)
         + \Lb 1 - i\,\frac{M^2_0\!-\!M^2}{\mu^2}\Rb G^{-}(q)
         \Bigg\}
       =\frac{1}{2\,\mu^2}
         \Big\{(1+i\,m_0) G^{+}(\kappa)
              +(1-i\,m_0) G^{-}(\kappa)
         \Big\}\nn
\eea
where we use notations:
\beq \label{VAR}
   \kappa \,=\,\frac{q^2}{\mu^2},\quad
   \kappa'\,=\,\frac{q'^2}{\mu^2},\quad
         E\,=\,- \frac{\omega}{\bas},\quad
      \bas\,=\,\frac{\as N_c}{\pi},\quad
         m\,=\,\frac{M^2}{\mu^2},\quad
       m_0\,=\,\frac{M^2_0-M^2}{\mu^2}\,.
\eeq

\subsection{The BFKL equation in momentum representation.}

The BFKL equation for Gribov-Zwanziger gluon propagator has been derived in our previous
paper\cite{GOLEM}, using the procedure that has been described in Ref.\cite{LLS}.

It has two parts: the gluon reggeization and the emission of gluons. The first one has
a general form\cite{BFKL}:
\beq \label{GGLTR}
  \omega_G(q)\,=\,G^{-1}(q)\,\Sigma(q)\quad\mbox{with}\quad
  \Sigma(q)\,=\,\int\frac{d^2 q'}{4\,\pi} G\Lb\vec{q}'\Rb\,G\Lb\vec{q} - \vec{q}'\Rb
\eeq
where $G(q)$ is given by \eq{GGLPR1}. The analytical expression for \eq{GGLTR} we will
discuss below (see also appendix A of Ref.\cite{GOLEM}).

\begin{figure}[ht]
\centering
\leavevmode
  \includegraphics[width=16cm]{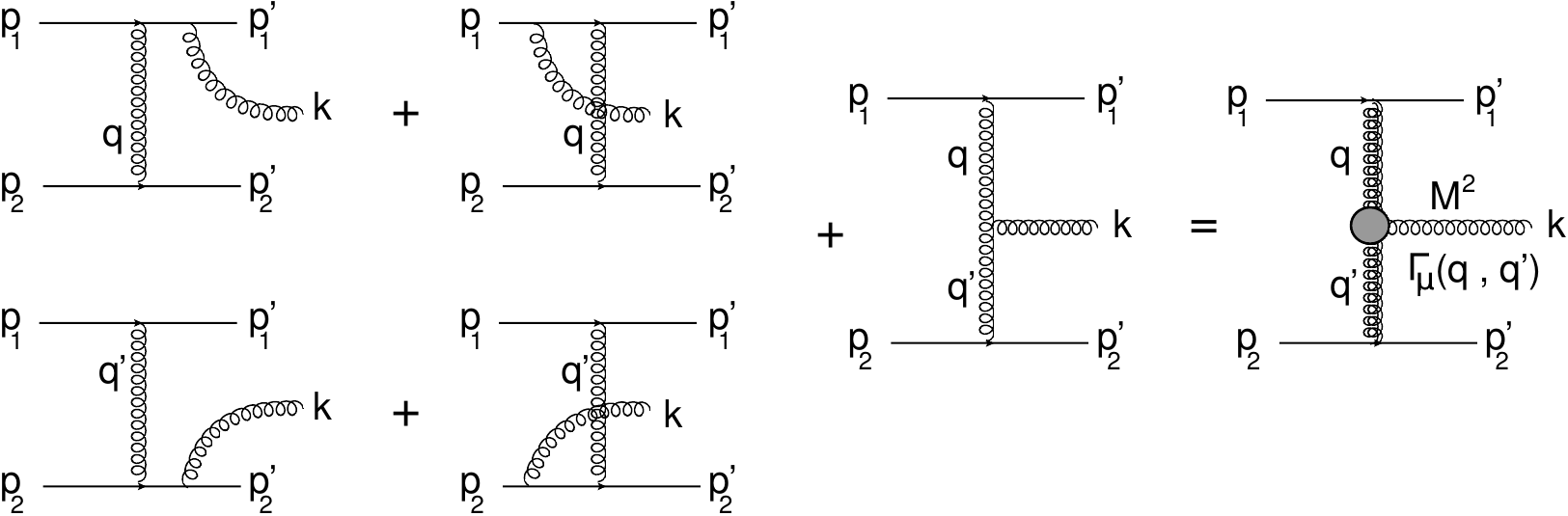}
  \caption{The first Feynman diagrams with gluon emission, whose sum leads
  to $\Gamma_\nu(q,q')$ (Lipatov vertex is denoted by the gray blob).}
\label{3}
\end{figure}

The emission kernel has been calculated in Ref.\cite{GOLEM} using the decomposition of
\eq{GLPR}. Indeed, using this decomposition we can treat the production of the gluon
as the sum of two sets of the diagrams (see \fig{3}) with $\tilde{M}^2 =\,\,i\mu^2$ and
with $\tilde{M}^2 =-\,i\mu^2$.

We sum the first diagrams of the gluon emission shown in \fig{3} to find the vertex
$\Gamma_\mu(q,q')$ for the kernel of the BFKL equation. It is easy to see that the sum
shown in \fig{3}, leads to the Lipatov vertex that has the following form\cite{LLS}:
\beq \label{V}
 \Gamma_\mu\Lb q, q'\Rb\, = \,- q^\perp_{\mu}\,-\,q'^\perp_{\mu}
     \,+\,p_{1,\mu} \Lb - G^{-1}(q) \frac{1}{p_1 \cdot k}\,+\,\frac{p_2\cdot k}{p_1\cdot p_2}\Rb
     \,-\,p_{2,\mu} \Lb - G^{-1}(q')\frac{1}{p_2 \cdot k}\,+\,\frac{p_1\cdot k}{p_1\cdot p_2}\Rb
\eeq
where $p_{1,\mu}$ and $p_{2, \mu}$ are the momenta of incoming particles (see \fig{3} for all notations).

Using \eq{GGLTR} and \eq{V}, the BFKL equation for Gribov-Zwanziger confinement
takes the form (for $\vec{Q}_T = 0$\footnote{%
$\vec{Q}_T$ is the momentum transferred by the BFKL Pomeron, a conjugate variable
to the impact parameter.}):
\beq \label{BFKLMR}
  \omega \,\phi(\omega, q)\,=
    \,-\,2\,\omega_G(q)\,\phi(\omega,q)
    \,+\,\bas \int \frac{d^2 q'}{\pi} G\Lb\vec{q} - \vec{q}^{\,'}\Rb\,\,\phi(\omega,q')
\eeq

This equation looks similar to the BFKL equation for a massive gluon \cite{LLS} in the
non-abelian Yang-Mills theories with a Higgs particle, which is responsible for mass
generation. However, we do not have a contact term in \eq{BFKLMR}. As we have discussed
in \cite{GOLEM} the absence of a contact term in our equation is a direct indication
that Gribov-Zwanziger confinement does not lead to a massive gluon.

Assuming that $\phi(q)$ depends only on $|\vec{q}|$, we can integrate the emission kernel
over the angle and in terms of the variable of \eq{VAR}, \eq{BFKLMR} takes the form:
\begin{subequations}
\bea
  E \,\phi(\kappa)\,\,&=&\,\,\,
     \underbrace{T(\kappa)\,\phi(\kappa)}_{\mbox{kinetic energy}} \,-\,
     \underbrace{\int d \kappa' \,K(\kappa,\kappa')\,\phi(\kappa')}_{\mbox{emission kernel}}
     \label{BFKLF1}\\
  &=& -\, T(\kappa)\,\phi(\kappa)\,\,-\,\,\int d\kappa' \,K(\kappa,\kappa')\,
      \Bigg\{\phi\Lb \kappa'\Rb \,-\,\frac{G(\kappa')}{G(\kappa)}\,\phi(\kappa)\Bigg\}
     \label{BFKLF2}
\eea
\end{subequations}
where
\begin{subequations}
\bea
  T(\kappa) \,\,&=&\,\,\frac{1}{4}G^{-1}(\kappa)\Big\{
      \,{\rm Re}
        \left(     \mzi^2 \,I_1(\mpi,\kappa)\right)
        \,+\,(1\,+\,m_0^2)\,I_2(m,\kappa)
    \Big\}\,,
  \label{GZ1}\\
  I_1(\mpi,\kappa)\,\,&=&\,\,
     \frac{2}{\sqrt{\kappa (\kappa + 4\,\mpi)}}
     \ln\left(\frac{\sqrt{\kappa}+\sqrt{\kappa+4\,\mpi}}{-\sqrt{\kappa}+\sqrt{\kappa+4\,\mpi}}\right)\,,
  \label{I1}\\
  I_2(m,\kappa)\,\,&=&\,\,
     -\frac{1}{\sqrt{4 m \kappa+\kappa^2-4}}
     \ln\Lb\frac{\kappa+2\,m\,-\,\sqrt{4 m \kappa+\kappa^2-4}}{\kappa+2\,m\,+\,\sqrt{4 m \kappa+\kappa^2-4}}\Rb\,,
  \label{I2}\\
  K(\kappa,\kappa')\,\,&=&\,\,
    {\rm Re}\left\{\frac{\mzi}{\sqrt{2\,\mpi\,(\kappa\,+\,\kappa')+\mpi^2+(\kappa\,-\,\kappa')^2}}\right\}\,,
  \label{GZ2}\\
  G(\kappa)\,\,&=&\,\,\frac{\kappa\,+\,m\,+\,m_0}{\Lb\kappa\,+\,m\Rb^2\,+\,1}\,,
  \label{GZ3}\\
  \mpi\,\,&=&\,\,m+i\,,\quad
  \mzi\,\, = \,\,1+i\,m_0\,.
  \label{MPI}
\eea
\end{subequations}

In \eq{BFKLF1} - \eq{GZ3} we use the variables which are given in \eq{VAR}.

\section{The basics of the spectrum for the master equation }

\subsection{ The equation for the eigenfunctions of the massless BFKL equation}

As has been mentioned the eigenfunctions of the massless BFKL equation
\beq \label{LKAP}
  \phi_{\mbox{\tiny BFKL}}(\kappa;\gamma)\,=\,\kappa^{\gamma - 1}\quad\mbox{with}\quad\gamma\,=\,\h\,+\,i\,\nu
\eeq
form the complete and orthogonal set of functions. Hence, we can expect that the solution to the master
equation can be written as the sum over these functions. By this reason we find instructive to consider
how the emission kernel of our master equation (see \eq{BFKLF1}) acts on the eigenfunctions of \eq{LKAP}:
\begin{subequations}
\bea
  \int \!\!d \kappa' K(\kappa,\kappa')\,\kappa'^{\,\ga-1}\!-\!\chi(\gamma)\,\kappa^{\gamma-1}\!\!&=&\!\!
     \int^\infty_0 \!\!\!d\kappa'\,
     {\rm Re}\,\left\{\frac{\mzi}{\sqrt{2\,\mpi\,(\kappa+\kappa')+\mpi^2+(\kappa-\kappa')^2}}\right\}
     \,\kappa'^{\,\gamma-1} -\chi(\gamma)\,\kappa^{\gamma-1}
  \label{SC1}\\
  &\to&\kappa^{\gamma-1} \Bigg\{\int_0^1\!\!d t\,
    \left(t^{\gamma-1}\!-1\right)
    {\rm Re}\!\left(\!-\frac{1}{\sqrt{(1-t)^2}}+\frac{\mzi}{\sqrt{(\mpi/\kappa)^2 + 2(t+1)\mpi/\kappa+(1-t)^2}}\right)
    \nn\\
  &+&\int_0^1\!\!d t \left(t^{-\gamma}\!-1\right)
    {\rm Re}\!\left(\!-\frac{1}{\sqrt{(1-t)^2}}+\frac{\mzi}{\sqrt{(t\mpi/\kappa)^2 + 2(t+1)t\mpi/\kappa+(1-t)^2}}\right)
    \!\!\Bigg\}
  \label{SC11}\\
  &\equiv&\kappa^{\gamma - 1} \,P(\kappa,\gamma)
  \label{SC12}
\eea
\end{subequations}
where the kernel of massless BFKL $\chi\Lb \gamma\Rb$ has the form \cite{BFKL}:
\beq \label{CHI}
  \chi\Lb \gamma\Rb \,\,=\,\,\psi(1-\gamma)\,+\,\psi(\gamma)-2\,\psi(1)
                    \,\,=\,\,\psi\Lb \h + i\,\nu\Rb
                        \,+\,\psi\Lb \h - i\,\nu\Rb \,-\,2\,\psi(1)
\eeq
where $\psi(z)$ is the Euler $\psi$-function (formula {\bf 8.36} of Ref.\cite{RY}).

In \eq{SC11} the region of integration over $\kappa'$ is divided in two: $\kappa' \,\leq \,\kappa$
and $\kappa'\,\geq \,\kappa$. In the first region the new variable is introduced $t = \kappa'/\kappa$,
while in the second the new variable is $t =\kappa/\kappa'$. In this way we have both $t$'s in the
region $(0,1)$. In addition, we subtracted in \eq{SC11} (terms with $1$ in the numerators of the
equation) the contribution from the Regge trajectory (see \eq{GGLTR}).

\begin{figure}[ht]
  \begin{tabular}{c c }
    \includegraphics[width=0.45\textwidth]{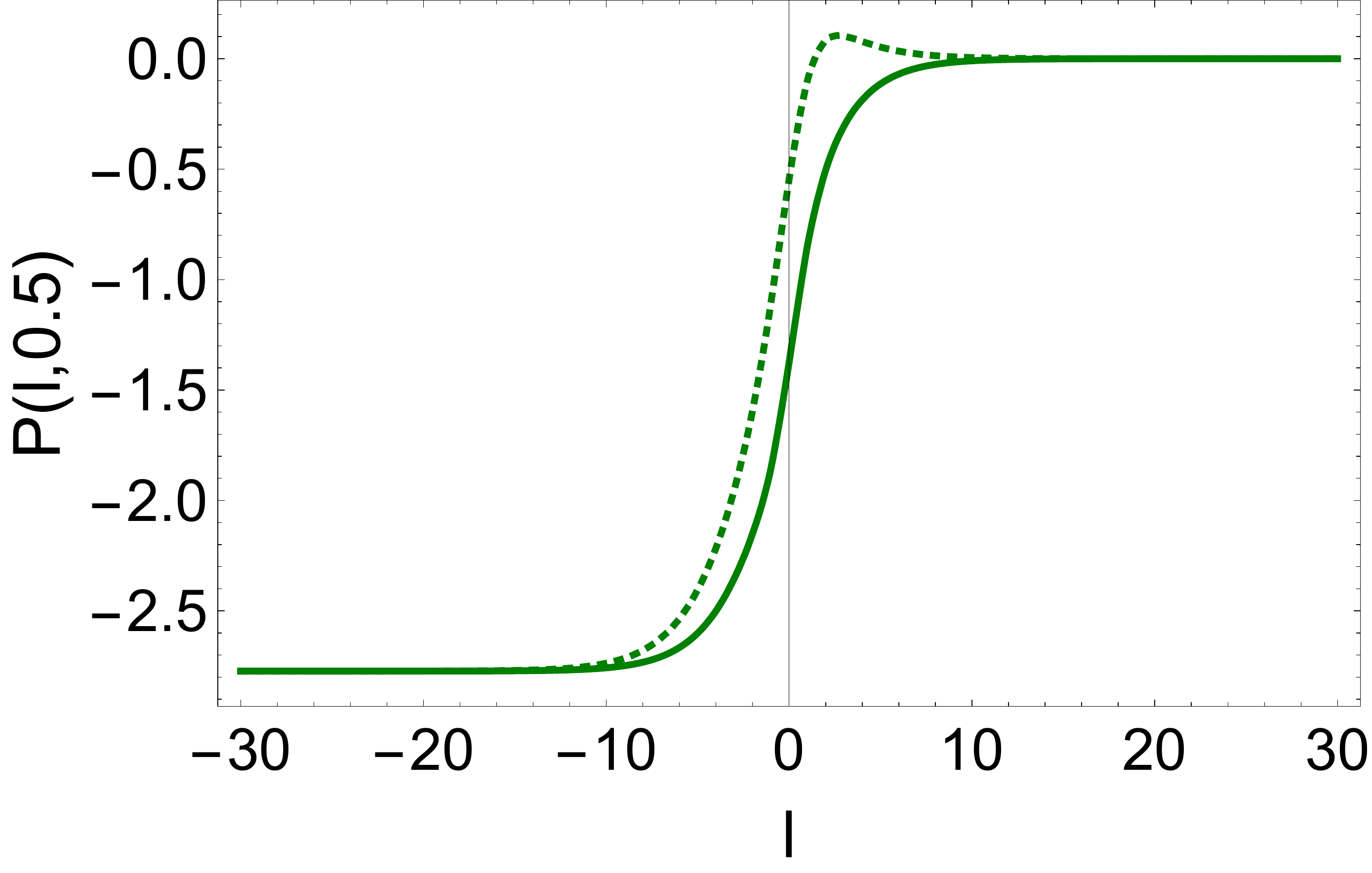} &
    \includegraphics[width=0.44\textwidth]{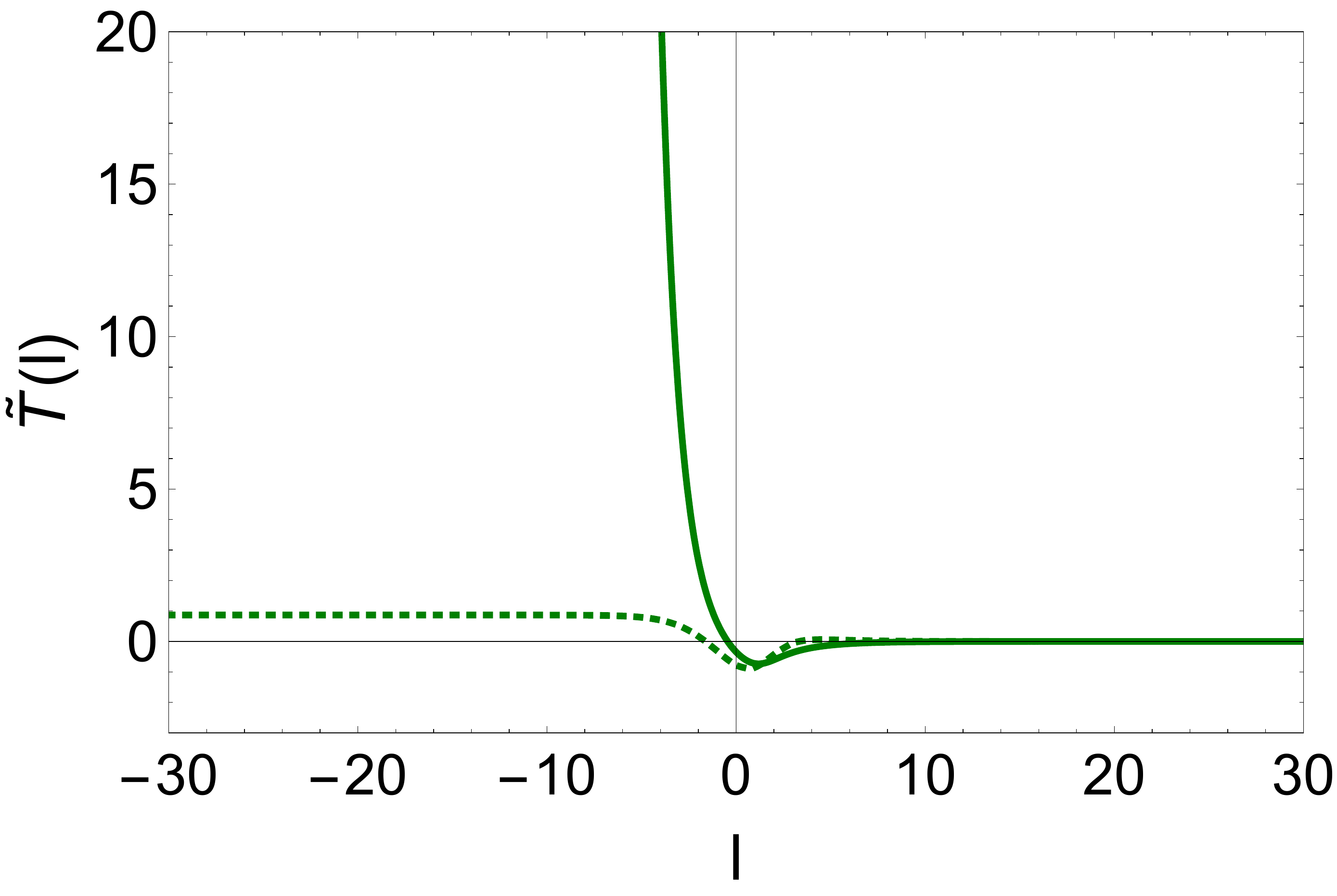} \\
    \fig{p}-a & \fig{p}-b\\
  \end{tabular}
  \caption{Functions $P\Lb\kappa=e^l,\gamma=0.5\Rb$ (\protect\fig{p}-a) and $\widetilde{T}\Lb e^l\Rb$
    (\protect\fig{p}-b) versus $l$. In these figures the solid lines describe the case $m = m_0 = 0$
    while the dotted ones correspond to $m =1.27, m_0 = 3.76$, which follows from the lattice QCD
    estimates \cite{DOS}.}
  \label{p}
\end{figure}

Using formula {\bf 3.211} of Ref.\cite{RY} we can express these integral over $t$ through
the Appel $F_1$ function (see Ref.\cite{RY} formulae {\bf 9.180-9.184}):
\bea \label{APEK}
  &&P(\kappa,\gamma)\,+\,\chi(\gamma)\,=\,
    {\rm Re}\Bigg\{\mzi\Bigg[
      \frac{\kappa}{\gamma\Lb \kappa+\mpi\Rb}
      \,\,F_1\!\left(\gamma;\h,\h;\gamma+1;
      \frac{\kappa}{\kappa - \mpi - 2\sqrt{-\kappa\mpi}},
      \frac{\kappa}{\kappa - \mpi + 2\sqrt{-\kappa\mpi}}\right)
    \\
  &&-\frac{\sqrt{4\kappa^3+\mpi(3\kappa+\mpi)^2}}{(\gamma-1) (\kappa+\mpi) \sqrt{4\kappa+\mpi}}
     \,\,F_1\!\left(1-\gamma;\h,\h;2-\gamma;
      \frac{(\kappa+\mpi)^2}{\kappa-\mpi - 2\sqrt{-\kappa\mpi}},
      \frac{(\kappa+\mpi)^2}{\kappa-\mpi + 2\sqrt{-\kappa\mpi}}\right)
    \nn\\
  &&+\,\ln\left(\frac{2}{1+\sqrt{1+4\kappa/\mpi}}\right)
    -\,\frac{\kappa}{\kappa+\mpi}\ln\left(
          \frac{1}{2\kappa}\left[
            \frac{\kappa+\mpi}{\mpi}\,\sqrt{\mpi(4\kappa+\mpi)} + 3\kappa+\mpi
          \right]
        \right)
    \Bigg]\Bigg\}
    \nn
\eea

From \eq{SC11} and \eq{APEK} (see also \fig{p}-a) we can see that $P(\kappa,\gamma)$ is rather
small and decreases at large positive $l = \ln\kappa$.

Since in \eq{SC11} we subtract the reggeization term we have re-defined the kinetic term in \eq{BFKLF1},
subtracting from $T\Lb \kappa\Rb$ of \eq{GZ1} function $L\Lb \kappa\Rb$ which is equal
\bea
  L(\kappa)&=&
        \int_0^1\! dt\,t\,\frac{\mzi}{\sqrt{\left(t\,\mzi/\kappa\right)^2 + 2 (t+1)\,t\,\mzi/\kappa + (1-t)^2}}
   \,+\,\int_0^1\! dt     \frac{\mzi}{\sqrt{\left(   \mzi/\kappa\right)^2 + 2 (t+1)   \,\mzi/\kappa + (1-t)^2}}
   \nn\\
  &=&\,\mzi\ln\left(\frac{1+\sqrt{1+4\kappa/\mpi}}{2}\right)
   +\,\frac{\kappa\,\mzi}{\kappa+\mpi}\ln\left(
        \frac{1}{2\kappa}\left[
           \frac{\kappa+\mpi}{\mpi}\sqrt{\mpi(4\kappa+\mpi)}+3\kappa+\mpi
        \right]
     \right)
\label{LL}
\eea
We denote
\beq \label{TT}
  \widetilde{T}(l)\,=\,T\Lb e^l\Rb\,-\,{\rm Re}\,L\Lb e^l\Rb
\eeq
$\widetilde{T}(l)$ is plotted in \fig{p}-b.

Using function $P\Lb l,\gamma\Rb$ and $\tilde{T}(l)$ we see that our equation for the function
$e^{(\ga - 1) l}$ has the form
\beq \label{SC5}
  \big(E  + \chi(\gamma)\big)\,
     e^{(\ga - 1) l}\,=\,
     e^{(\ga - 1) l} \big(P(l,\gamma) \,- \,\tilde{T}(l)\big)\,=\,
     e^{(\ga - 1) l}\,\widetilde{P}(l,\gamma)
\eeq

\begin{figure}[ht]
  \centering
  \begin{tabular}{c c}
    \includegraphics[width=0.5\textwidth]{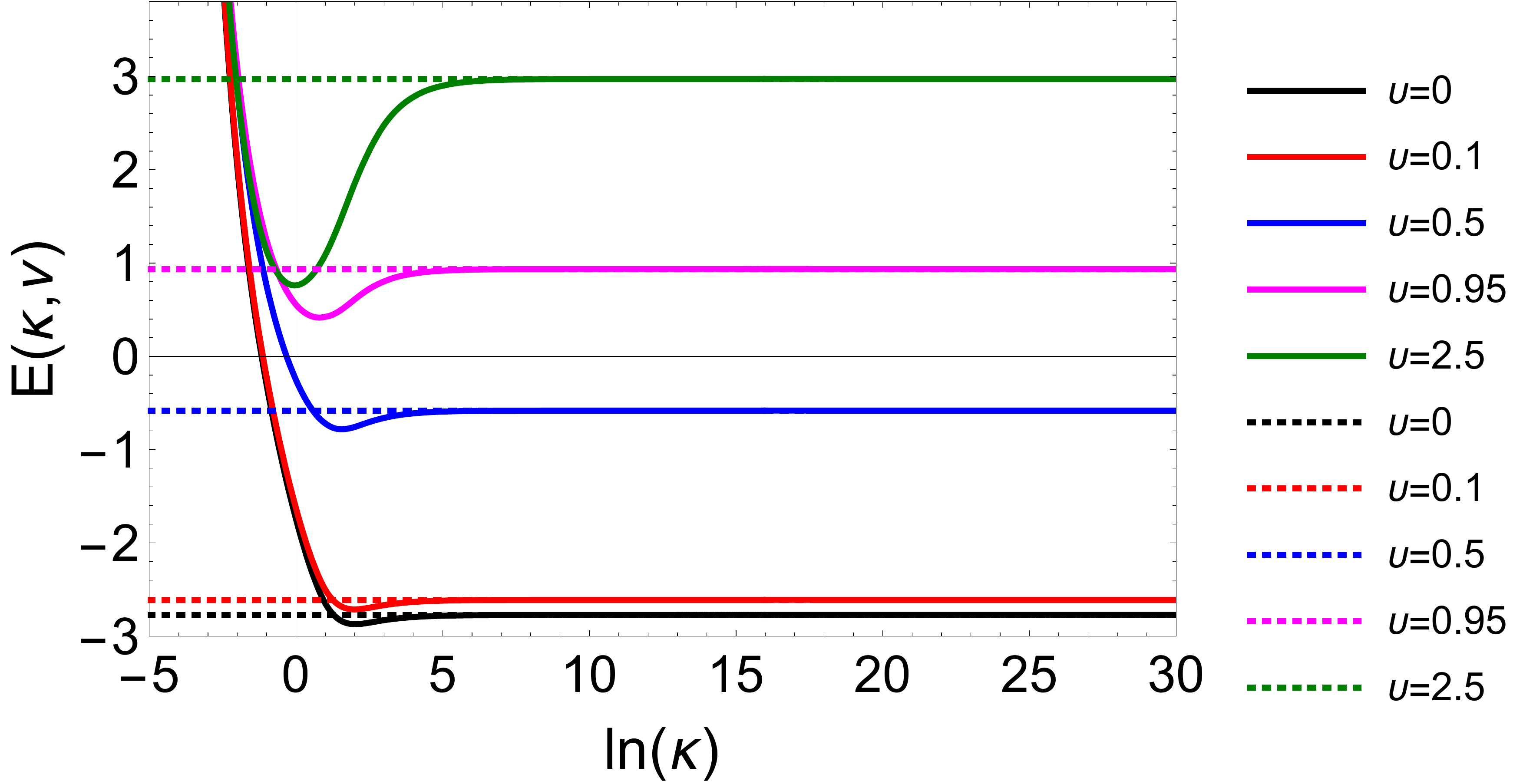} &
    \includegraphics[width=0.5\textwidth]{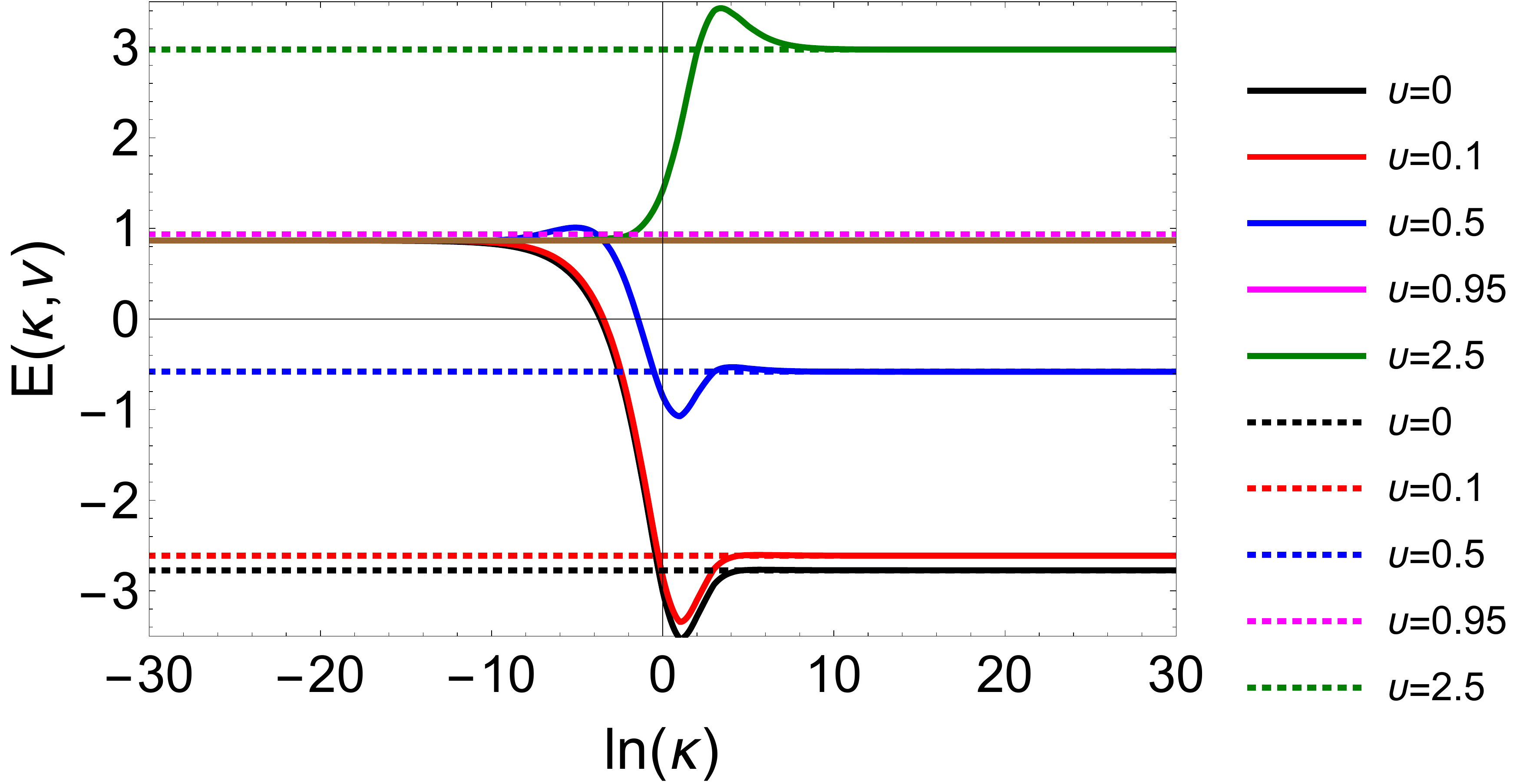} \\
    \fig{omdifnu}-a &\fig{omdifnu}-b\\
  \end{tabular}
  \caption{The dependence of the energies of the master equation (see \eq{BFKLF1}) versus $l=\ln\kappa$
    for the eigenfunctions of \eq{LKAP} with different values of $\gamma \h\,+\,i \nu$. The dotted lines
    describe the eigenvalues of the massless BFKL equation. \fig{omdifnu}-a describes the energies for
    the Gribov gluon propagator (see \eq{GLPR}) while \fig{omdifnu}-b corresponds to the gluon propagator
    of \eq{GGLPR1} with $m=1.27$ and $m_0=3.76$ which follows from the lattice QCD estimates \cite{DOS}.
    The brown line in \fig{omdifnu}-b shows $E= T(\kappa=0)$.}
\label{omdifnu}
\end{figure}

In \fig{omdifnu} we plot
\beq \label{OMDIFNU}
  E(\nu)\,=\,- \chi(\nu) \,+\,\widetilde{P}(l,\nu)
\eeq
fixing $\gamma\,=\,\h\,+\,i\,\nu$. \fig{omdifnu}-a gives \eq{OMDIFNU} for $m = m_0 = 0$, which
corresponds to the Gribov gluon propagator while in \fig{omdifnu}-b the energy is plotted for
the lattice QCD gluon propagator with $m = 1.27$ and $m_0 = 3.76$ \cite{DOS}.

One can see that the wave functions of the massless BFKL show up as the eigenfunctions of the
master equation in the kinematic region of large $l =\ln\kappa \,\gg\,1$. Generally speaking,
it means that the eigenvalues of \eq{BFKLF1} could be (1) the same as the massless one or
(2) could be selected out due to behaviour at small $\kappa$, leading to the set of the
eigenvalues, which is more restricted than the massless BFKL one. In addition, of course,
could be some discrete states, whose wave functions decrease steeper than $\kappa^{\gamma-1}$
at large values of $\kappa$. From the numerical solution (see below) we see that there is no
selection and all energies of the massless BFKL equations occur as the eigenvalues of the
master equation. We can understand this, since the BFKL eigenvalues of \eq{LKAP} are twice
degenerate. One can see this since the eigenvalues of the massless BFKL equation do not
depend on the sign of $\nu$. At $l < 0$ we expect the the eigenfunction of the massive
BFKL equation should be constant. Replacing this behaviour by the boundary condition
$\phi(l = 0; \nu) = \mbox{Const}$ we see that we can satisfy this boundary condition choosing
$\phi(l)\,=\,C_1\,\phi_{\mbox{\tiny BFKL}}(l,\nu)
           + C_2\,\phi_{\mbox{\tiny BFKL}}(l,\nu)
        \,=\,\sin\Lb l\,\nu + \varphi\Rb$ where $\varphi$ is the phase. One can see that
we do not bring any selection with this procedure.

However, the eigenfunctions of the massless BFKL equation appear the eigenfunctions of \eq{SC5}
also for $\kappa \,\ll\,1$ ($l \,=\,\ln \kappa\,\ll\,-1$) (see \fig{omdifnu}-b) but only for
the case of $m\neq 0$ and $m_0 \neq 0$ with the eigenvalue $E_0 \,=\,T(\kappa=0)\,=\,0.866$
for any value of $\nu$(brown line in \fig{omdifnu}-b). The independence of $\nu$ means that
the eigenvalue $\omega_0$ is infinitely degenerate. \fig{p}-c shows that in \eq{OMDIFNU}
$E(\nu)\,=\,\big\{-\chi(\nu)\,+\,P(l,\ga)\big\}\,-\,\tilde{T}(l)$ the term in $\big\{\dots\big\}$
vanishes al $l<0$ (see \fig{p}-c), while $\widetilde{T}(l)$ approaches a constant (see \fig{p}-b).

In principle, such solutions could be rejected for the master equation if the behaviour at small
$\kappa$ cannot be matched with the behaviour at large $\kappa$. However, it looks very unlikely.
Indeed, any function of the following type: $\phi(\kappa)\,=\,P_n(2\,\kappa - 1)\,\Theta(1-\kappa)$
where $P_n(z)$ is the Legendre polynomial (see Ref.\cite{RY} formulae {\bf 8.91}), is orthogonal
to $\phi(\kappa) = \mbox{Const}$ at $l<0$ (for $n>1$) and satisfies \eq{SC5}. The numerical
calculations, which we will discuss below, confirm that $E = E_0$ appears as the eigenvalue
of the generalization of the BFKL equation (see \eq{BFKLF1}).

\subsection{General features of the spectrum}
Following the general pattern of Ref.\cite{LLS} we can re-write \eq{BFKLF1} in the coordinate space, introducing
\beq \label{GF1}
  \Psi(r)\,\,=\,\,\int \frac{d^2 q_T}{(2\,\pi)^2} e^{i \vec{r}\cdot\vec{q}_T } \,\phi\Lb q_T\Rb
\eeq
The equation takes the following form
\beq \label{HGR}
  E\,\Psi(r)\,\,=\,\,{\cal H}\,\Psi(r)
\eeq
with
\beq \label{HGR1}
  {\cal H}\,\,=\,\,\underbrace{T\Lb\hat{\kappa}\Rb}_{\mbox{\small kinetic energy}}
          \,\,-\,\,\underbrace{U\Lb r \Rb}_{\mbox{\small potential energy}}
          \,\,=\,\,T\Lb\hat{\kappa}\Rb\,-\,G(r)
\eeq
where $\hat\kappa=-\nabla^2_r$ is the momentum operator and $G(r)$ is equal to
\beq \label{GF2}
  G(r)\,=\,\int \frac{d^2 q_T}{(2\,\pi)^2} e^{i \vec{r}\cdot\vec{q}_T }\,G\Lb q_T\Rb
        =\,\mzi K_0\Lb\sqrt{\mpi}\,r\Rb+\mzi^* K_0\Lb\sqrt{\mmi}\,r\Rb
        \,\,\xrightarrow{r\,\gg\,m}
        \,\,{\rm Re}\left\{\frac{\mzi}{\mpi^{1/4}}\sqrt{\frac{\pi}{2\,r}} e^{-\sqrt{\mpi}r}\right\}
\eeq

For large $r$, $G(r)$ exponentially decreases as one can see from \eq{GF2}. Hence,
at large $r$ \eq{HGR} takes the following form:
\beq \label{HGR2}
  E\Psi(r)\,\,=\,\,T\Lb\hat{\kappa}\Rb\,\Psi(r)
\eeq
with the eigenfunctions that have the following form
\beq \label{GF3}
  \phi\Lb \vec{r}\Rb    \,\,\sim\,\, e^{ i\sqrt{\kappa^2} r}, ~~~\kappa^2\,>\,0;
  \quad\phi\Lb\vec{r}\Rb\,\,\sim\,\, e^{-\sqrt{-\kappa^2} r}, ~~~\kappa^2\,<\,0.
\eeq

\begin{figure}[ht]
  \begin{tabular}{cc}
    \includegraphics[width=0.5\textwidth]{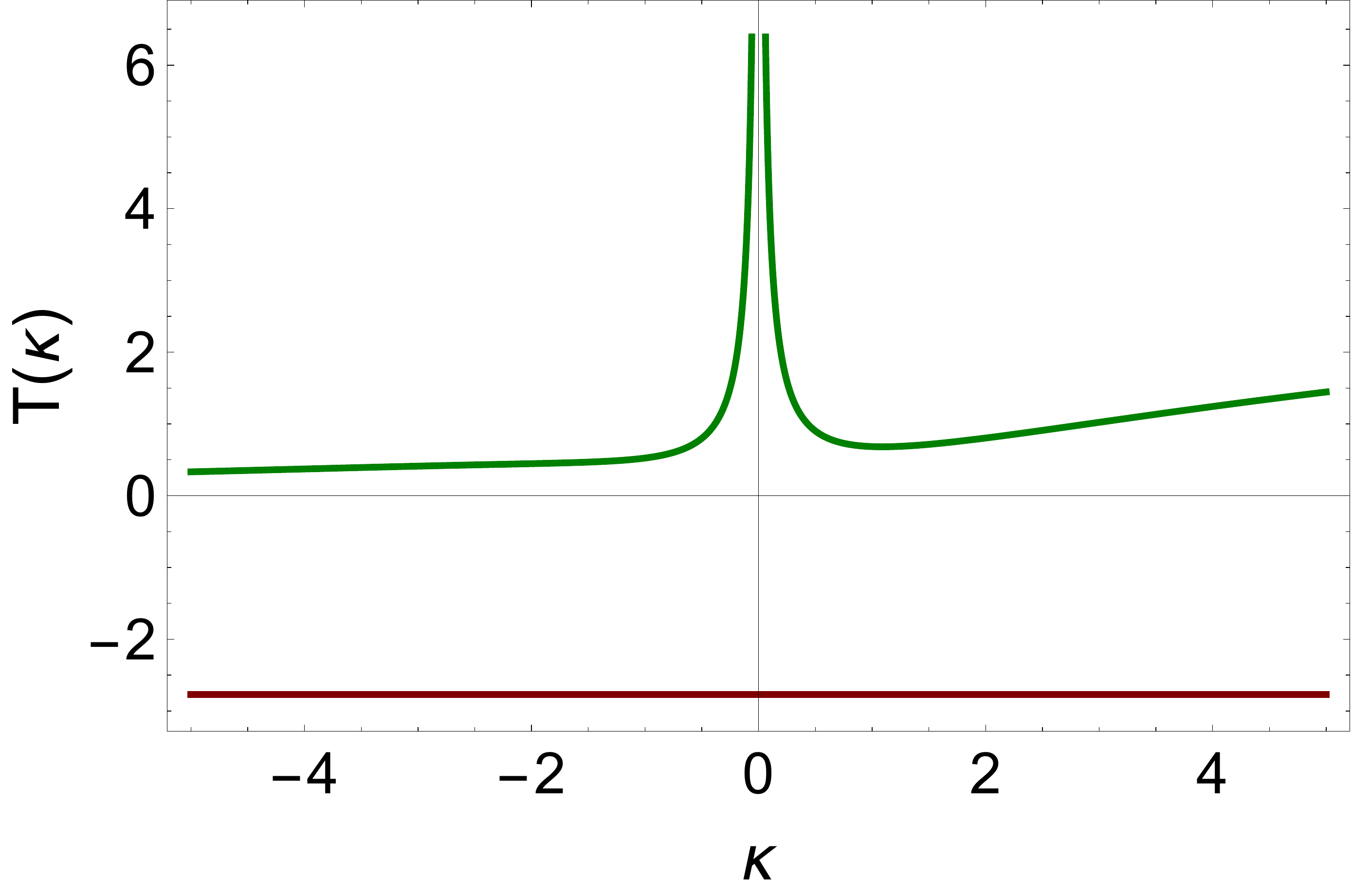} &
    \includegraphics[width=0.5\textwidth]{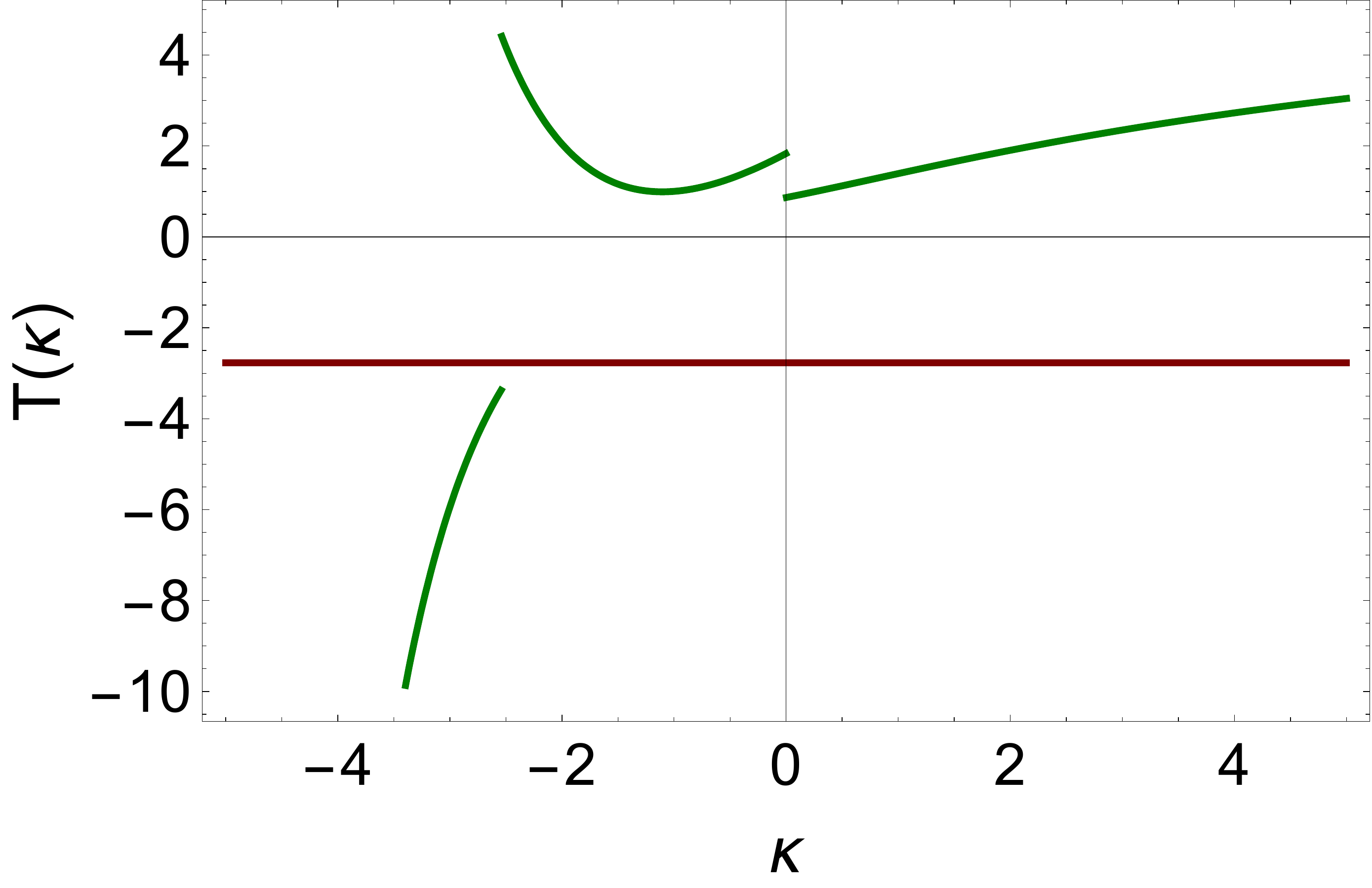}
  \end{tabular}
  \caption{ $T(\kappa)$ versus $\kappa$ for Gribov's propagator of \eq{GLPR} ($m=m_0=0$) and for
    the gluon propagator that describes the lattice QCD estimates ($m = 1.27$ and $m_0 =3.76$).
    The red line shows $E=-4\ln 2$, which is the ground state for the massless BFKL equation.}
\label{t}
\end{figure}

Denoting the large asymptotic behaviour of the eigenfunction as
$\Psi(r) \xrightarrow{r\,\gg\,\,1/\mu} \,\,\exp\Lb-\sqrt{a}\,r\Rb$,
we see that the energy is equal to
\beq \label{HGR3}
  E \,\,=\,\, T(-a)
\eeq
On the other hand, in the region of small $r$ \eq{HGR1} reduces to the massless QCD BFKL equation
(see the previous section and \cite{BFKL,LIP}:
\beq \label{HGR4}
  E\,\Psi\Lb r \Rb \,\Psi\Lb r \Rb \,\,=\,\,{\cal H}_0 \,\Psi\Lb r \Rb
\eeq
where \cite{LIP}
\beq \label{H0}
  {\cal H}_0\,=\,\ln p^2 \,+\,\ln |r|^2\,-\,2\psi(1)
\eeq
The eigenfunctions of \eq{HGR4} are $\Psi(r)\,=\,r^{2(1-\gamma)}$, and the eigenvalues of \eq{HGR4}
can be parametrized as a function of $\gamma$ (see \eq{CHI}). Therefore, for $r\,\to\,0$ we have
the eigenvalue which is equal to
\beq \label{HGR5}
  E\,\,=\,\,\chi(\gamma)
\eeq
From \eq{HGR3} and \eq{HGR5} we can conclude, that the value of $a$ and $\gamma$ are correlated, since
\beq \label{HGR6}
  E\,\,=\,\,\chi(\gamma)\,\,=\,\,T(-\,a)
\eeq
Based on \eq{HGR5} (see also the previous section) we expect that the minimum eigenvalue is equal
to $\chi(\h) \,=\,-\,4\,\ln 2$. From \fig{t} we can see that \eq{HGR6} is violated. For Gribov's
propagator of \eq{GLPR} $T(\kappa)$ is positive for all values of $-\infty\,<\,\kappa\,<\,+\infty$.
For the gluon propagator that describes the lattice QCD estimates ($m = 1.27$ and $m_0 =3.76$
we can see from \fig{t}-b that $T(\kappa)$ is negative at $\kappa\,<\,0$, and therefore, \eq{HGR6}
can be satisfied.

On the other hand the estimates of \eq{HGR3} contradict the result of Ref.\cite{GOLEM} that
the eigenfunction of the master equation with the Gribov's gluon propagator exhibit the
power-like decrease at long distances. We believe that a resolution of this inconsistency
is intimately related to the definition of $\Psi(r)$. In particular, instead of \eq{GF1}
we suggest to introduce the following transform to the coordinate space. First we introduce
a new $\tilde{\phi}(\kappa)\,\,=\,\,G^{-1}(\kappa)\,\phi(\kappa)$. For this function
\eq{BFKLMR} takes the form:
\beq \label{BFKLN}
   \omega\,\tilde{\phi}(\omega,q)\,\,=
      \,\,-\,2\omega_G(q)\,\tilde{\phi}(\omega,q)
      \,\,+\,\,\bas \int \frac{d^2 q'}{\pi}
        \left\{G^{-1}(q)\,G\Lb\vec{q}-\vec{q}^{\,'}\Rb\,\,G\Lb q'\Rb\right\}
        \,\tilde{\phi}\Lb\omega,q'\Rb
\eeq
The eigenfunction in the coordinate space has the form:
\beq \label{GF4}
  \Psi(r)\,\,=\,\,\int \frac{d^2 q_T}{(2\,\pi)^2} e^{i\vec{r}\cdot\vec{q}_T}\,\tilde{\phi}\Lb q_T\Rb
\eeq
and the master equation has the form of \eq{HGR} with the potential energy, which has the different form:
\beq \label{GF05}
  {\cal H}\Psi(r)\,\,=\,\,\underbrace{T\Lb\hat{\kappa}\Rb}_{\mbox{\small kinetic energy}}\,\Psi(r)
    \,\,-\,\,\int d^2\, r'\underbrace{U\Lb\vec{r},\vec{r}'\Rb}_{\mbox{\small potential energy}}\,\Psi\Lb r'\Rb
\eeq
with
\bea \label{GF5}
  &&U\Lb r,r'\Rb\,\,=\,\,
    \int \frac{d^2 q_T }{(2\,\pi)^2} e^{i\vec{r}\cdot\vec{q}_T }
    \int \frac{d^2 q'_T}{(2\,\pi)^2} e^{i\vec{r}'\cdot\vec{q}'_T }
    \left\{G^{-1}(q)\,G\Lb\vec{q}-\vec{q}'\Rb\,\,G\Lb q'\Rb\right\}\\
  &&=\,\,\int d^2\,r'' \,K_0\big((m+m_0)\,|\,\vec{r}\,-\,\vec{r}''\,|\,\big)
    \left(\Lb-\nabla^2_{r''}+m\Rb^2\,+\,1\right)\,
    G\Lb r''\Rb\,G\Lb\vec{r}''\,-\,\vec{r}'\Rb\nn
\eea
One can see that for $m = m_0 = 0$ the potential energy $U(r,r')\propto\,ln\Lb r\Rb$ and \eq{HGR2}
turns out to be incorrect. For $m\,\neq\,0$ and $m_0\,\neq\,0$ the potential energy decreases
exponentially at long distances. Hence, in this case \eq{HGR2} holds.

\subsection{Eigenfunctions in the vicinity of $E_0 = T(\kappa=0)$}

As we have discussed above, there is a possibility, that the master equation has the eigenvalues
in addition to the eigenvalues of the QCD BFKL equation. These states should have the wave functions
that decrease much steeper that the eigenfunction of \eq{LKAP}. From \fig{omdifnu} one can expect
that the vicinity of $E = E_0 = T(\kappa=0)$ can provide such states. Indeed, in the vicinity
$E\,\to\,E_0$ $T(\kappa)$ takes the form
\bea
  T(\kappa)\,&=&{\rm Re}\Bigg\{
     \frac{\mzi\mpi\mmi}{4(m+m_0)}
       \left[\frac{\mzi}{\mpi} + \frac{i}{2} \mzi^* \ln\left(\frac{\mmi}{\mpi}\right)\right]
   +\,\frac{\kappa\,\mzi}{4}
    \Bigg[
      \left(1-\frac{\mzi\mzi^*}{(m+m_0)^2}\right)
      \left(\frac{\mzi}{\mpi}+\frac{i}{2} \mzi^* \ln\left(\frac{\mmi}{\mpi}\right)\right)
      \nn\\
   &+&\frac{\mpi\mmi}{m+m_0}
    \left\{
      -\frac{\mzi}{6\,\mpi^2}
      + \mzi^* \left(-\frac{1}{2}+\frac{i}{4} m \ln\left(\frac{\mmi}{\mpi}\right)\right)
    \right\}
  \Bigg] \Bigg\}
  \label{E01} \\
  &\equiv& E_0 \,\,+ \,\,E'_0\,\kappa \label{E02}
\eea

\eq{BFKLF1} takes the following form in vicinity of $\kappa \to 0$
\beq \label{E03}
  (E - E_0 - E'_0\,\kappa) \phi(\kappa)\,\,=
    \,\,-\int d\kappa'\,K\Lb\kappa=0,\,\kappa'\Rb\,\phi(\kappa')
    \,\,-\,\,\kappa\,\int d\kappa'\,\frac{\partial K\Lb \kappa,\kappa'\Rb}{\partial\kappa}\Bigg{|}_{\kappa=0}\phi\Lb\kappa'\Rb
    \,\,+\,\,{\cal O }\Lb\kappa^2\Rb
\eeq
Introducing
$$\epsilon = \frac{E-E_0}{E'_0\,-\,\int d\kappa'\,\frac{\partial K\Lb\kappa,\kappa'\Rb}{\partial\kappa}\Big{|}_{\kappa=0}}$$
one can see from \eq{E03} that $\phi\Lb \kappa\Rb$ has a singularity:
\beq \label{E04}
  \phi(\kappa)\Big{|}_{\kappa\,\to\,\epsilon}\,\,=\,\,\frac{\mbox{Const}}{\epsilon\,-\,\kappa}
\eeq
or, in other words the wave function has the form:
\beq \label{E05}
  \phi(\kappa)\,\,=\,\,\frac{\mbox{Const}}{\epsilon\,-\,\kappa}\,+\,\phi_{\,\rm bg}(\kappa)
\eeq
where $\phi_{\,\rm bg}$ is the function which has no singularities. Since $E=E_0$ is multiple
degenerate eigenvalue, the sum of functions of \eq{E05} is also an eigenfunction.

It is instructive to note that the eigenfunctions of \eq{E05} does not appear for the QCD BFKL
equation. As we have seen, the origin of such eigenfunctions is in the fact that typical
$\kappa'$ in \eq{E03} are about the values of mass and not equal to zero.

\subsection{Resume}

Concluding this section we wish to emphasize two results that we have proved. First, the eigenvalues
of the massless BFKL equation, generally speaking, are expected to be the eigenvalues of the master
equation. In principle, it is possible that the behaviour of the wave functions at small values of
$\kappa$ could select out some of the eigenvalues of the BFKL equation in QCD. However, due to double
degeneracy of each of the massless BFKL eigenvalues (see, that \eq{CHI} has symmetry $\nu \to -\nu$)
the boundary conditions at $\kappa \to 0$ does not lead to a loss of the eigenvalues of the master
equation in comparison with the massless BFKL equation.

Second, it is possible that the eigenvalues of the master equation have a reacher structure than
the eigenvalues of the BFKL equation in QCD. Indeed, could be states with the wave functions that
are suppressed at large $\kappa$: $\phi(\kappa) \,\,\ll\,\, \kappa^{-\h + i\,\nu}$. An example of
such function could be \eq{E04}. As we see from \fig{omdifnu} the eigenfunctions with $E\,=\,E_0$
have infinite degeneracy and all of them are eigenfunctions that have not been present in the massless
BFKL equation.

The separate problem is the state with the wave function that decreases steeper than the eigenfunction
of the massless BFKL equation but with the eigenvalue which is smaller than $E_{\rm min} = -4\ln 2$.

At the moment we cannot answer this question without finding the numerical solution to the master equation.

\section{Numerical solution}
\subsection{General approach}
Generally speaking we need to solve the equation which has the following structure:
\beq \label{NS1}
  E\,\phi\Lb \kappa\Rb\,\,=\,\int d \kappa' {\cal K}\Lb \kappa, \kappa'\Rb \,\phi\Lb \kappa'\Rb
\eeq
where ${\cal K}$ is defined in \eq{BFKLF2}. The advantage of using \eq{BFKLF2} in comparison
with \eq{BFKLF1}, have been discussed in Appendix B of Ref.\cite{GOLEM}.

For numerical solution we discretize the continuous variables $\kappa$ and $\kappa'$ using
the logarithmic grid $\{\kappa_n\}$ with $N+1$ nodes
\bea\label{NS2}
  \kappa_{n} &=& \kmin \exp \left(n \De_\kappa\right), \quad \De_\kappa=\frac{1}{N}\, \ln\left(\kmax/\kmin\right),
  \quad n=0,...,N,
\eea
where the values of $\kmin$ and $\kmax$ are fixed. In the most details we consider the case
with $\kmin=10^{-10}$, $\kmax=10^{65}$ and $N=2000$, but we investigated the dependence of
the solution on the values of $\kmin$, $\kmax$ and $N$.

In the discrete variables \eq{NS1} can be approximated in the form
\beq \label{NS3}
  E \phi\Lb \kappa_n\Rb\, \,=\,\, \sum^{N}_{m=0} \kappa_m\,\De_\kappa\, K\Lb \kappa_n, \kappa_m\Rb\, \phi\Lb\,\kappa_m\Rb
\eeq

Introducing the notations:
  $\phi\Lb \kappa_n\Rb \equiv \phi_n$ and
  $\kappa_m\,\De_\kappa\,{\cal K}\Lb \kappa_n, \kappa_m\Rb\equiv\, {\cal K}_{n m}$
we can re-write \eq{NS3} in the matrix form
\beq \label{NS4}
  E\,\phi_n\,\,=\,\, \sum^{N}_{m=0} {\cal K}_{n m}\,\phi_m\qquad\mbox{or}\qquad
  E\,\vec{\phi}\,\,=\,\,\,{\cal \mathbf K}\,\vec{\phi}
\eeq
where vector $\vec{\phi}$ has $N+1$ components $\phi_n$ and ${\cal\mathbf K}$ is $(N+1)\times(N+1)$
matrix. We need to find the roots of the characteristic polynomial $p\Lb E \Rb$ of the matrix
${\cal \mathbf K}\,-\,E\,{\mathbf I}$ where ${\mathbf I} $ is the identity matrix. Hence, we
need to solve the secular equation
\beq \label{NS5}
  p\Lb E\Rb\,\,=\,\,\mbox{det}\Lb{\cal \mathbf K}\,-\,E\, {\mathbf I}\Rb\,\,=\,\,0
\eeq

We solve \eq{NS5} for several equations. First, we found the solution to our new \eq{BFKLF2}
in two cases: for the Gribov propagator of \eq{GLPR} and for the propagator of \eq{GZ3}.
In the first case we need to put $m=0$ and $m_0=0$ in Eqs.~(\ref{GZ1})-(\ref{GZ3}) while
in the second we need to choose $m=1.27$ and $m_0=3.76$ in these equations. Such values
follow from the lattice estimates for the gluon propagator. Second, we solve the original
BFKL equation for QCD, which has the form:
\beq \label{BFKLQCD}
  E \,\phi_{\mbox{\tiny BFKL}}\Lb \kappa\Rb\,\,=
    \,\,-\,\int\frac{ d \kappa'}{|\kappa - \kappa'|}
      \Bigg( \phi_{\mbox{\tiny BFKL}}\Lb \kappa'\Rb
         \,\,-\,\,\frac{\kappa}{\kappa'}\phi_{\mbox{\tiny BFKL}}\Lb \kappa\Rb
      \Bigg)
    \,\,-\,\, \int \frac{ \kappa\,\,d \kappa' }{\kappa'\,\,\sqrt{\kappa^2 + 4 \kappa'^2}}\,\,\phi_{\mbox{\tiny BFKL}}\Lb \kappa\Rb
\eeq
We believe that we need to compare our numerical procedure, with the equation which
has the analytical solution, both to check the accuracy of our numerical estimates
and to evaluate our transition to the continuous limit. Recall, that the numerical
solution gives the discrete spectrum of the eigenvalues instead of the continuous one.
In addition we solve the equation, which was derived in Ref.\cite{LLS} for non-abelian
gauge theory with the Higgs mechanism for mass generation. This theory is not QCD, since
it has no confinement of quarks and gluons. However, it has the same colour structure as
QCD and introduces the dimensional scale: the mass of Higgs boson. Solving the BFKL
equation for this theory we could find out what is more essential: the new dimensional
scale or specifics related to the confinement. We will call below this approach ``the model''
and the main BFKL equation for this model takes the form:
\bea \label{BFKLHIGGS}
  \hspace{-0.5cm}&& E \phi\Lb \kappa\Rb\,=\,
     \underbrace{\frac{\kappa+1}{\sqrt{\kappa}\sqrt{\kappa+4}}
        \ln\frac{\sqrt{\kappa+4}+\sqrt{\kappa}}{\sqrt{\kappa+4}-\sqrt{\kappa}}\phi\Lb \kappa\Rb}_{\mbox{\small kinetic energy term}}
     \,-\,
     \underbrace{\int^{\infty}_{0}\-\frac{d \kappa' \phi\Lb \kappa'\Rb}{\sqrt{(\kappa-\kappa')^2\,+\,2 (\kappa + \kappa') + 1}}}_{\mbox{\small potential energy term}}
     \,+\,
     \underbrace{\frac{N^2_c + 1}{2 N^2_c}\frac{1}{\kappa + 1}\int^{\infty}_0 \frac{\phi\Lb \kappa'\Rb \,d \kappa'}{\kappa' + 1}}_{\mbox{\small contact term}}
\eea
This equation has been investigated in detail. In Ref.\cite{LLS} it was proven that
the solution of this equation coincide with the solution to the massless BFKL equation,
which is known analytically and can be used as a control of the accuracy of our numerical
calculations.

\subsection{Eigenvalues: the general characteristics}

\begin{figure}[ht]
  \begin{tabular}{cc}
    \includegraphics[width=0.5\textwidth]{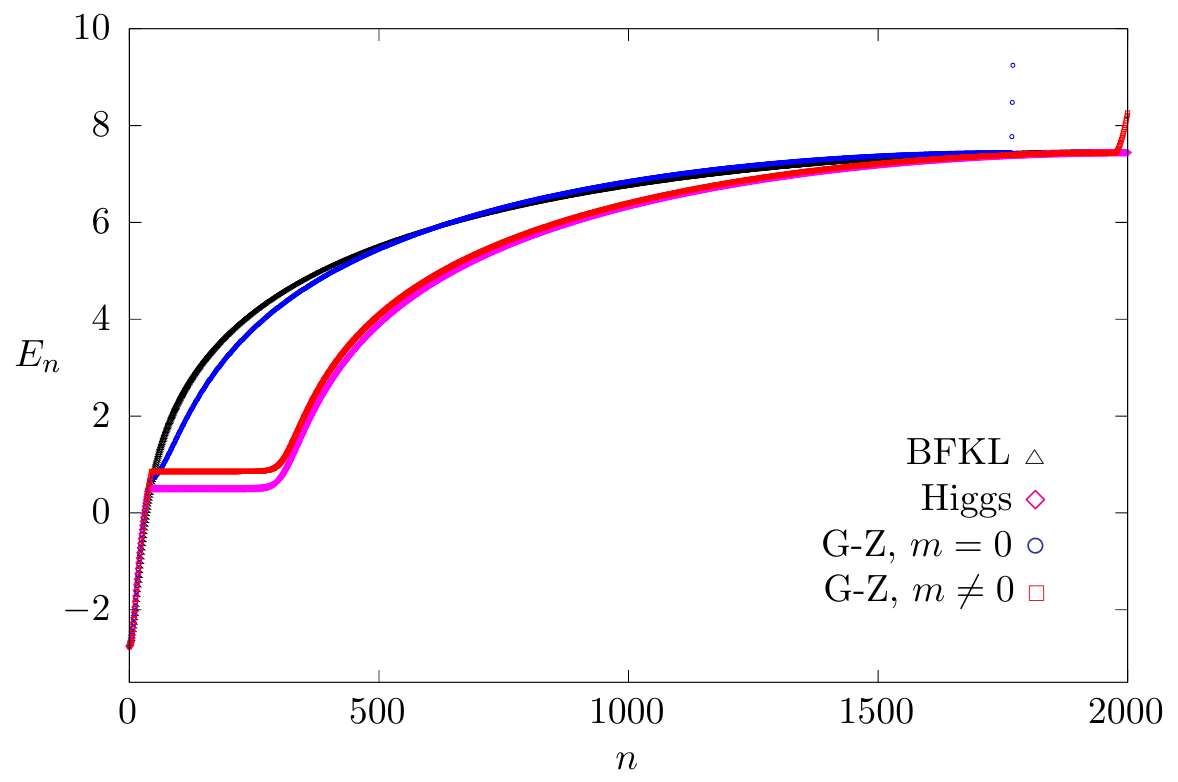} &
    \includegraphics[width=0.5\textwidth]{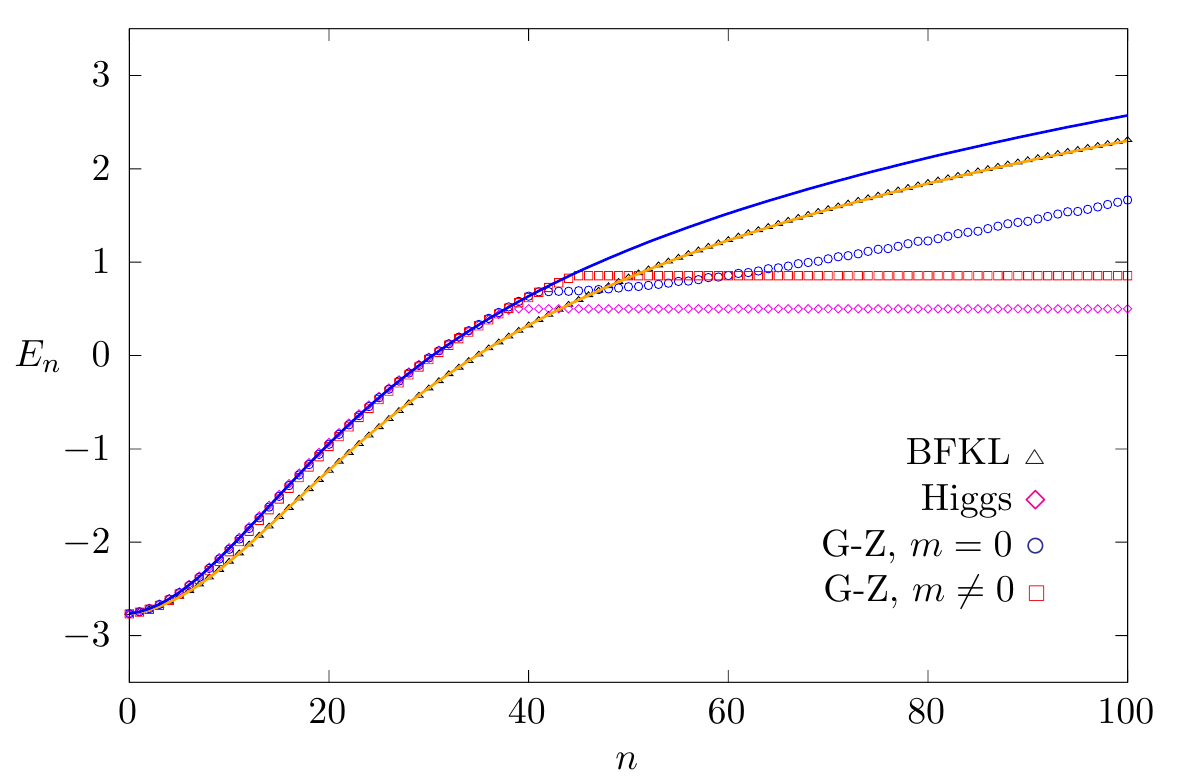} \\
    \fig{rootgen}-a &\fig{rootgen}-b\\
  \end{tabular}
  \caption{
     The eigenvalues $E_n$ of four equations: the BFKL equation in QCD (BFKL, see \eq{BFKLQCD}),
     the BFKL equation for the modl (Higgs, see \eq{BFKLHIGGS}), for \eq{BFKLF2} in the case of
     Gribov propagator (G-Z, $m=0$) and in the case of lattice QCD propagator of \eq{GGLPR1}
     (G-Z, $m\neq0$). The solid lines in \fig{rootgen}-b show the eigenvalues calculated using
     \eq{EVN0} with $\beta_n$, that were taken from the pattern of zeros of the eigenfunctions
     given by \eq{EF1}, \eq{EF2} and \eq{EF4}. All results correspond to solutions with
     $\kmin=10^{-10}$, $\kmax=10^{65}$ and grid size $N=2000$ (see \eq{NS2}).
  }
  \label{rootgen}
\end{figure}

The eigenvalues of these four equations are shown in \fig{rootgen}. One can see that (1) numerical
estimates do not show the discrete eigenvalues with energy $E \,\,<\,\,E_{\min} = -4 \ln 2$, where
$E_{\min}$ is the minimal energy of the massless BFKL equation; and (2) none of the eigenvalues of
the massless BFKL equation has been selected out in accord with our expectations in section III-A.

From \fig{rootgen} we see that all eigenvalues can be divided in three regions: the eigenvalues
$E_n \,\leq\,E_0= T\Lb \kappa = 0\Rb$, the multiply degenerate eigenvalue $E_0$ and $E_n\,\geq\,E_0$.

For $E_{\rm min} \,\leq E_n \,\leq E_0$ there are no other eigenvalues except the massless BFKL ones.
Indeed, we can describe these eigenvalues using the following formulae:
\bea \label{EVN0}
  E(n)     &=& -2\,\psi\Lb 1 \Rb \,+\, \psi\Lb \h \,+\,i \beta(n) \Rb \,-\, \psi\Lb \h\,- \,i \beta(n) \Rb\\
  \beta(n) &=& \,a_\beta\,(n \,+\,1), ~~~ a_\beta = c_\beta / \ln \Lb \kmax/m^2_\beta \Rb \nn
\eea

For the QCD BFKL equation $c_\beta=3.015$ and $m^2_\beta=\kmin$, while for all other three equations
we can put $c_\beta=3.140$ and $m^2_\beta=0.0042$ (see \fig{bvk}-a). Hence, \fig{rootgen}-b and Table~I
demonstrate a new phenomenon: the eigenvalues of all three equations, which introduce a dimensional
scale in the BFKL approach, turns out to be the same. Actually, they are the eigenvalues of the QCD BFKL
equation, as it shows \eq{EVN0} and \fig{rootgen}-b. In \fig{rootgen} the eigenvalues from \eq{EVN0}
are shown by the solid lines and one can see that all these values for $E_n \leq E_0$ can be perfectly
described by this equation. \eq{EVN0} can be interpreted as an indication that the transition to the
continuous limit reduces to replacement
  $\h + i \beta(n) \,\,\to\,\,\h \,+\,i\,\nu\,\equiv \,\gamma$.
In these new variables the eigenvalues of the QCD BFKL looks familiar\cite{KOLEB}:
\beq \label{BFKLQCDE}
  E\Lb \gamma \Rb\,\,=\, -2\, \psi\Lb 1 \Rb \,+\,\psi\Lb\gamma\Rb \,-\,\psi\Lb 1 - \gamma\Rb
\eeq

Table~I, in which we put the first 20 roots of the secular equation, illustrates these points.
First we see that solution to the QCD BFKL equation gives the eigenvalues, which are quite close
to the analytical estimates (see \eq{EVN0}). This indicates that our method of numerical solving
provides a good accuracy. As we can see, in both cases the lowest eigenvalue becomes quite close
to $E_{\min} = - 4 \ln2$ and difference is negligibly small (of the order of $5\times10^{-3}$ for
$\kmin=10^{-10}$ and $\kmax=10^{65}$).

\begin{figure}[ht]
\begin{tabular}{cc}
  \includegraphics[width=0.5\textwidth]{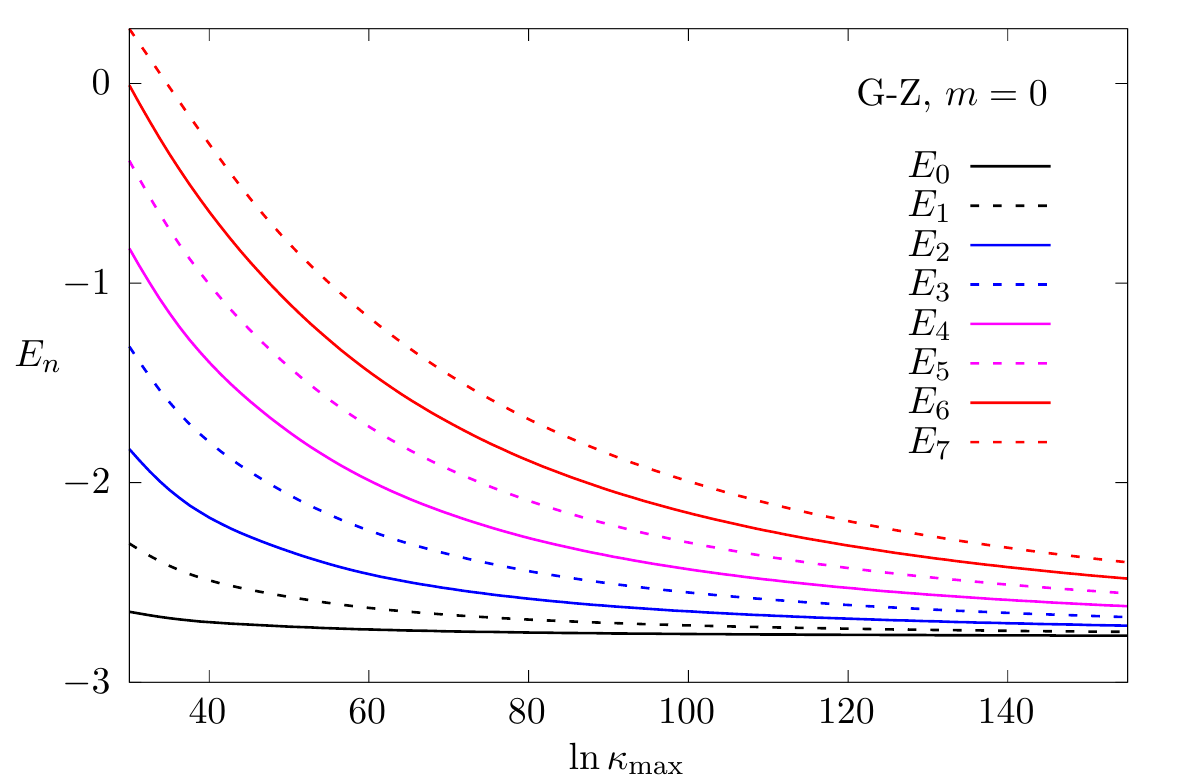} &
  \includegraphics[width=0.5\textwidth]{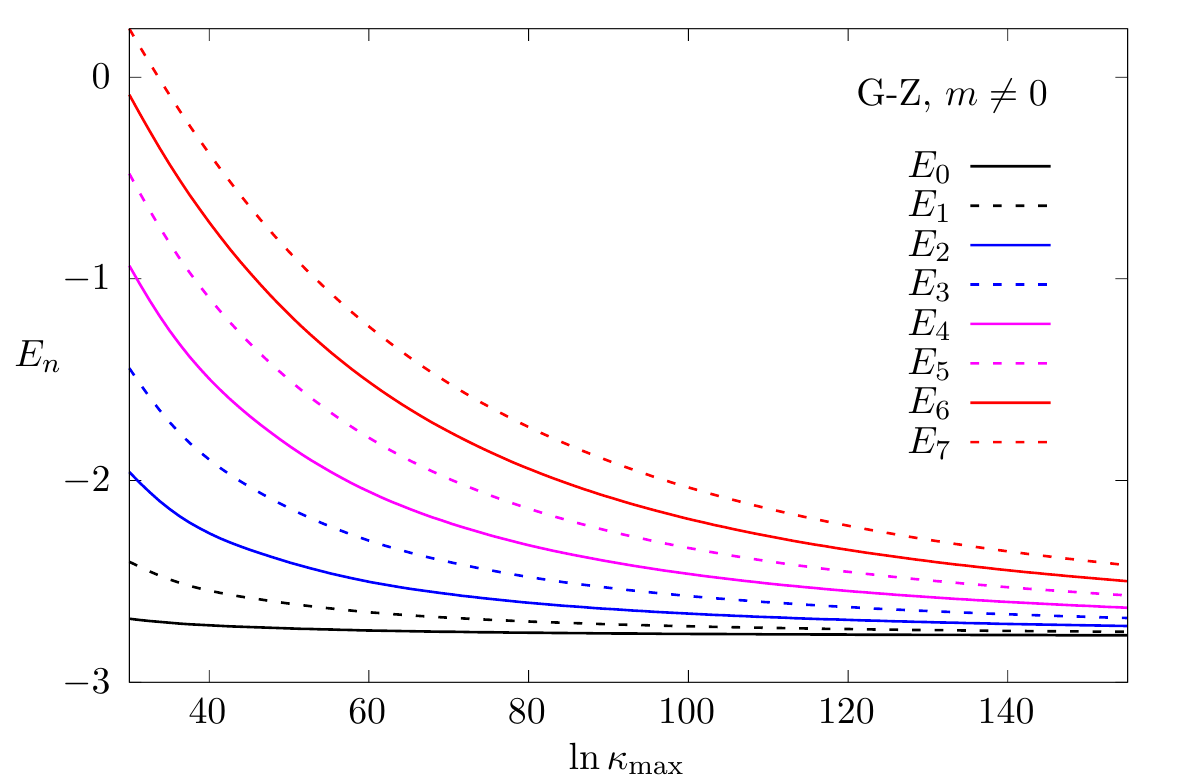} \\
  \fig{root}-a & \fig{root}-b\\
\end{tabular}
\caption{
  The first several eigenvalues of \eq{BFKLF1} versus $\ln\Lb\kmax\Rb$.
  \fig{root}-a for the Gribov propagator of \eq{GLPR} and \fig{root}-b
  for the lattice propagator of \eq{GGLPR1}.
}
\label{root}
\end{figure}

\fig{root} shows the dependence of the first 7 roots versus the value of $\kmax$. One can see that
when $\kmax$ grows ($\kmax\to\infty$) the distance between neighboring roots decreases rapidly,
inferring the smooth transition to the continuous limit.

As we can see from \fig{rootgen} at $E_n = E_0 = T\Lb \kappa = 0\Rb$ for three equations, that
introduce a new dimensional scale, we have multiple degenerate eigenvalue. For the Gribov propagator
this degeneration is not very large but in other cases it is so large that we can expect something
like Bose-Einstein condensation at this energy. The general structure of the eigenfunction at this
value of energy we have discussed in section III-C and will consider below. For the scattering
amplitude all these eigenfunctions correspond to the cross sections that decrease as a power of
energy and, because of this, do not show up in the high energy scattering processes.

\begin{center}
\begin{table}
\begin{tabular}{||r|c|c|c|c|c||}
\hline\hline
n & ~~~$E_n$ (QCD)~~~ & ~~~$E_n$ (Higgs)~~~ & ~$E_n$ (G-Z, $m=0$)~ & ~$E_n$ (G-Z, $m\neq0$)~ & ~~$E_n$ (\eq{EVN0})~~ \\
\hline\hline
 0 & -2.7675 & -2.7657 & -2.7660 & -2.7666 & -2.7657  \\ \hline
 1 & -2.7519 & -2.7448 & -2.7457 & -2.7483 & -2.7452  \\ \hline
 2 & -2.7261 & -2.7103 & -2.7123 & -2.7178 & -2.7114  \\ \hline
 3 & -2.6905 & -2.6630 & -2.6665 & -2.6753 & -2.6650  \\ \hline
 4 & -2.6456 & -2.6036 & -2.6088 & -2.6211 & -2.6067  \\ \hline
 5 & -2.5919 & -2.5332 & -2.5403 & -2.5561 & -2.5377  \\ \hline
 6 & -2.5301 & -2.4529 & -2.4620 & -2.4810 & -2.4588  \\ \hline
 7 & -2.4610 & -2.3640 & -2.3751 & -2.3968 & -2.3715  \\ \hline
 8 & -2.3854 & -2.2677 & -2.2808 & -2.3048 & -2.2768  \\ \hline
 9 & -2.3040 & -2.1653 & -2.1802 & -2.2060 & -2.1760  \\ \hline
10 & -2.2177 & -2.0581 & -2.0746 & -2.1018 & -2.0705  \\ \hline
11 & -2.1273 & -1.9472 & -1.9651 & -1.9932 & -1.9612  \\ \hline
12 & -2.0336 & -1.8337 & -1.8528 & -1.8815 & -1.8492  \\ \hline
13 & -1.9373 & -1.7186 & -1.7387 & -1.7675 & -1.7356  \\ \hline
14 & -1.8391 & -1.6029 & -1.6236 & -1.6523 & -1.6212  \\ \hline
15 & -1.7397 & -1.4872 & -1.5083 & -1.5368 & -1.5068  \\ \hline
16 & -1.6396 & -1.3724 & -1.3936 & -1.4215 & -1.3930  \\ \hline
17 & -1.5394 & -1.2588 & -1.2800 & -1.3072 & -1.2804  \\ \hline
18 & -1.4395 & -1.1470 & -1.1680 & -1.1944 & -1.1695  \\ \hline
19 & -1.3403 & -1.0374 & -1.0579 & -1.0835 & -1.0605  \\ \hline
20 & -1.2421 & -0.9303 & -0.9501 & -0.9748 & -0.9539  \\ \hline
\hline
\end{tabular}
\caption{
  The first 20 eigenvalues $E_n$ for BFKL equation in QCD (QCD, see \eq{BFKLQCD}),
  for the model, developed in Ref.\cite{LLS} (Higgs, see \eq{BFKLHIGGS}), for \eq{BFKLF2}
  in the case of Gribov propagator (G-Z $m=0$) and in the case of lattice QCD propagator
  of \eq{GGLPR1} (G-Z, $m \neq 0$). Last column contains values defined by \eq{EVN0}.
}
\label{t1}
\end{table}
\end{center}

\subsection{Eigenfunctions for $E\,\,\leq\,\,E_0\,\,=\,T\Lb \kappa = 0\Rb$}

For the QCD BFKL equation(see \eq{BFKLQCD}) the eigenfunctions are given by \eq{LKAP}
and for the numerical solutions they take the form:
\beq \label{EF1}
  \phi^{\mbox{\tiny BFKL}}_n\Lb \kappa\Rb\,\,=\,\,\frac{\alpha_n}{\sqrt{\kappa}}\sin\Lb \beta_n \,\ln \kappa\,\,+\,\,\varphi_n\Rb
\eeq
The eigenfunction for \eq{BFKLHIGGS} have been discussed in Ref.\cite{LLS} and can be described as follows:
\beq \label{EF2}
  \phi_n \Lb \kappa\Rb\,\,=\,\,\frac{1}{\sqrt{\kappa\,+\,4}}\sin\Lb \beta_n \, \mbox{Ln}(\kappa)\,\,+\,\,\varphi_n\Rb
\eeq
where
\beq \label{EF3}
  \mbox{Ln}(\kappa)\,\,=\,\,
  \frac{\sqrt{\kappa\,+\,4} + \sqrt{\kappa}}{\sqrt{\kappa\,+\,4} - \sqrt{\kappa}}\,\,\xrightarrow{\kappa\,\gg\,1}\,\,\ln (\kappa)
\eeq

\begin{figure}[hb]
\begin{tabular}{cc}
  \includegraphics[width=0.5\textwidth]{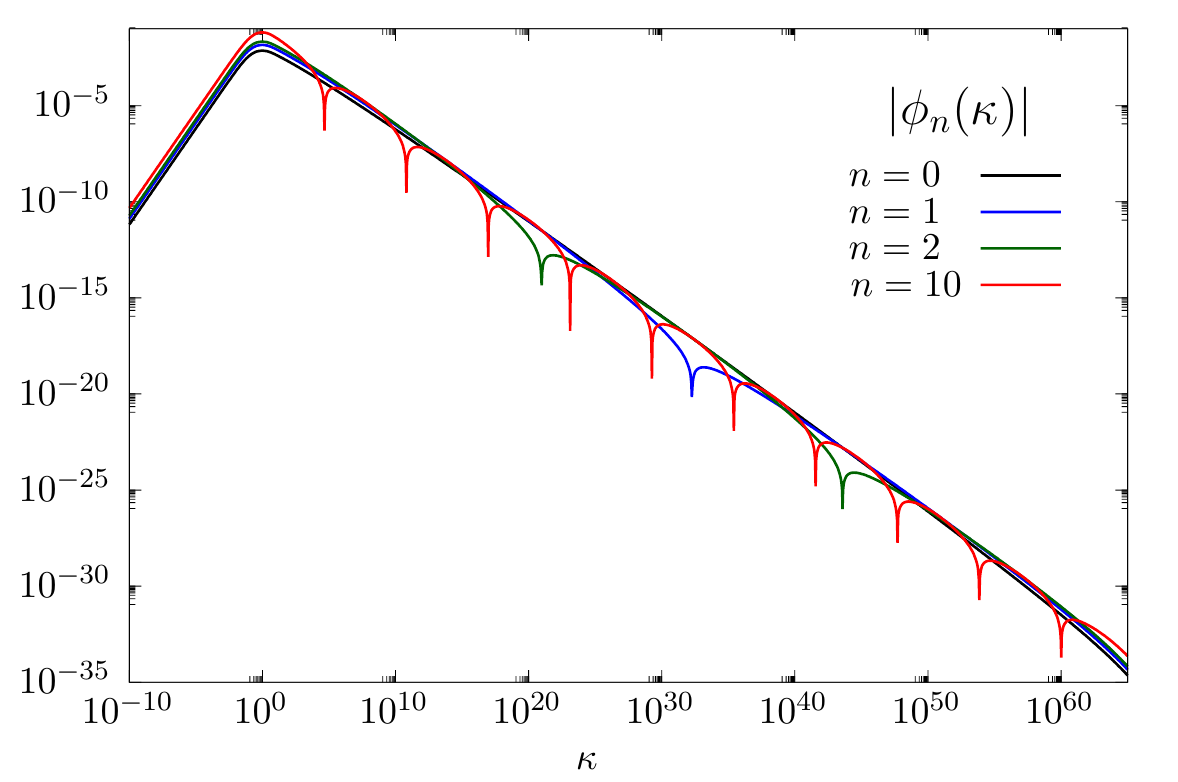} &
  \includegraphics[width=0.5\textwidth]{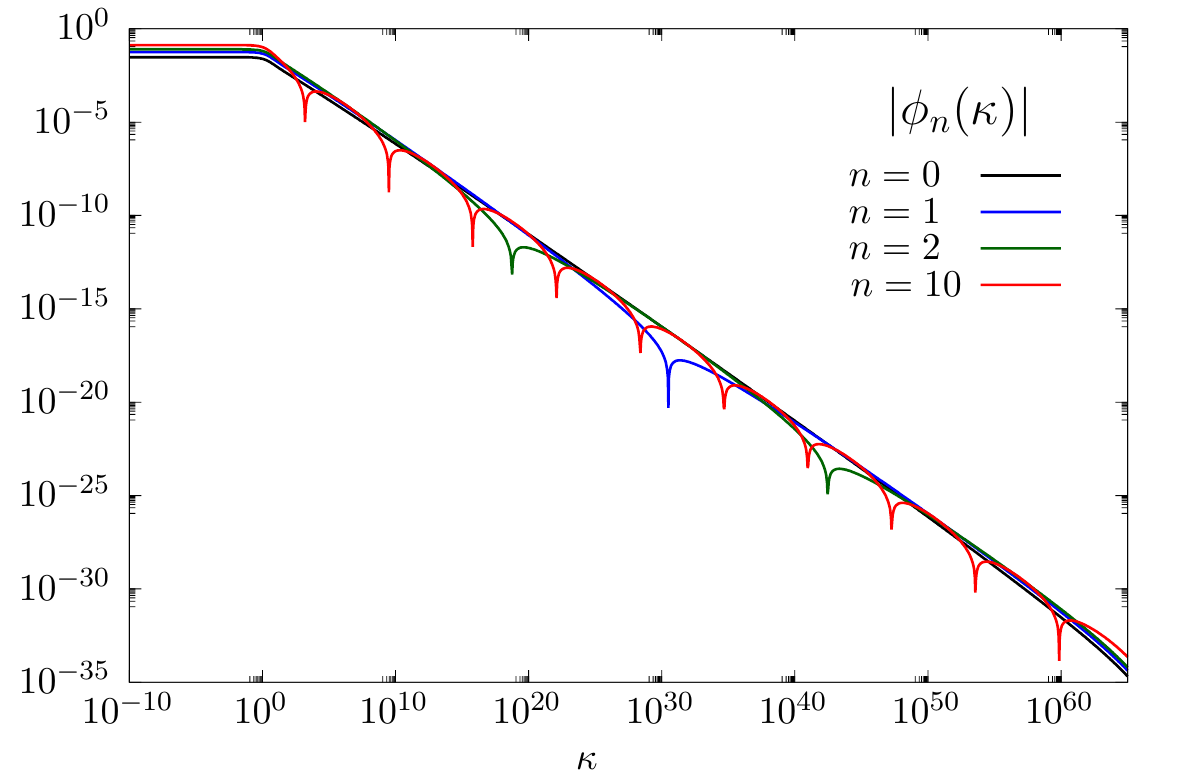} \\
  \fig{ef}-a &\fig{ef}-b\\
\end{tabular}
\caption{
  The examples of the eigenfunctions of \eq{BFKLF2}. \fig{ef}-a for the Gribov propagator of \eq{GLPR},
  i.e. $m=0,m_0=0$ in \eq{GZ3}. \fig{ef}-b for the lattice QCD propagator of \eq{GZ3} with $m=1.27,m_0=3.76$\,.
  Functions $\phi_n(\kappa)$ for $n=0,1,2$ and $10$. One an see that every $\phi_n(\kappa)$ has $n$-zeros.
}
\label{ef}
\end{figure}

Several examples of the eigenfunctions for \eq{BFKLF2} are shown in \fig{ef}. One can see that
the number of zeros follows the usual pattern of a quantum mechanical approach: the minimum
energy state has no zeros. The next has one and so on. At large $\kappa\,$
$\phi_n\Lb \kappa\Rb\,\propto\,\sin\Lb \alpha_\beta\,n \,\ln \kappa\Rb$ or in other words
$\phi_n\Lb \kappa\Rb = C_1 \phi_{\mbox{\tiny BFKL}}\Lb \kappa; \h + i \beta_n\Rb \,
                    +\,C_2 \phi_{\mbox{\tiny BFKL}}\Lb \kappa; \h - i \beta_n\Rb,$
where $\phi_{\mbox{\tiny BFKL}}\Lb \kappa; \gamma\Rb$ are given by \eq{LKAP}.

For $\kappa \,\geq 1$ all eigenfunctions can be parameterized in the following way:
\beq \label{EF4}
  \phi_n\Lb \kappa\Rb\,\,=\,\,\frac{\alpha_n\,(\kappa+m)}{\sqrt{(\kappa\,+\,a_n)^3}}\,\sin\Lb \beta_n Ln(\kappa) \,+\, \varphi_n\Rb
\eeq
where
\beq \label{EF5}
  Ln(\kappa) \,\,=\,\, \frac{\kappa}{4}
  \Big\{
    \mbox{Re}\Lb \mzi^2 \,I_1 \Rb
    \,+\,        \mzi \mzi^*  \,\,I_2
  \Big\}
  \,\,\xrightarrow{\kappa \,\gg\,1} \,\, \ln \kappa
\eeq

\begin{figure}[ht]
\centering
\begin{tabular}{c c}
  \includegraphics[width=0.5\textwidth]{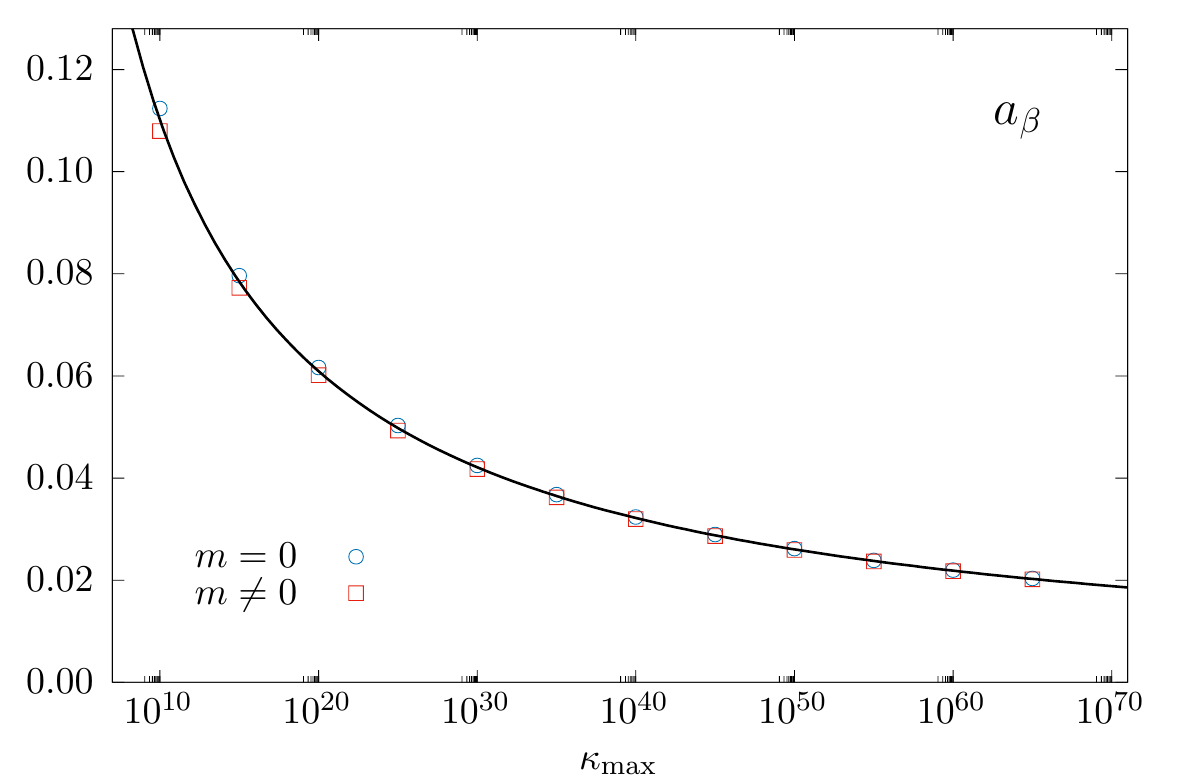}  &
  \includegraphics[width=0.5\textwidth]{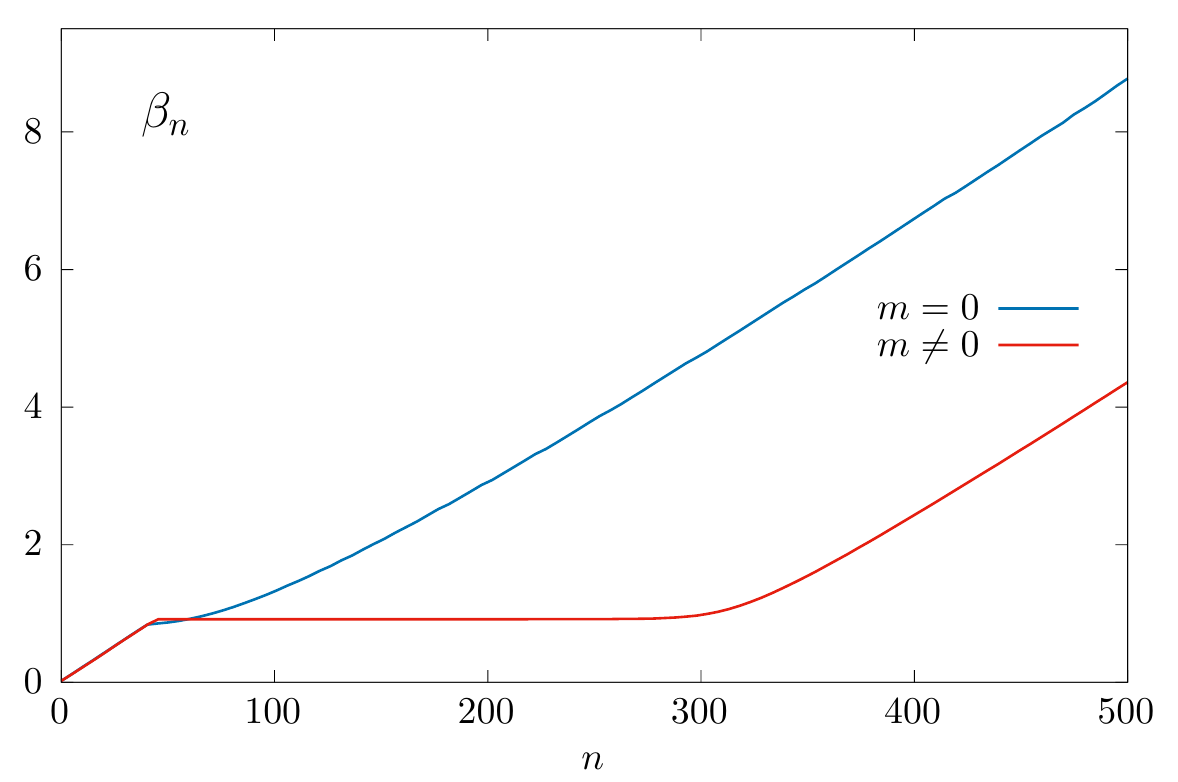} \\
  \fig{bvk}-a & \fig{bvk}-b\\
\end{tabular}
\caption{
  \fig{bvk}-a: $a_\beta$ versus $\kmax$ for Gribov's and lattice QCD gluon propagators.
  The solid line corresponds to \eq{EF6} with $m^2_\beta = 0.0042$ for both cases.
  \fig{bvk}-b: $\beta_n$ of \eq{EF4}  for the eigenfunctions $\phi_n(\kappa)$ obtained
  with $\kmin=10^{-10}$ and $\kmax=10^{65}$ at $\kappa \,>\, 1$ versus $n$.
}
\label{bvk}
\end{figure}

Parameter $\beta_n$ has the simple form defined by \eq{EVN0}
\beq \label{EF6}
  \beta_n \,\,=\,\,a_\beta\,(n+1), \quad a_\beta = \frac{3.140}{\ln(\kappa_{\rm max}/m^2_\beta)}
\eeq
It was found that for the Gribov's propagator ($m = 0$, $m_0 = 0$) and for the propagator
with $m \neq 0$, $m_0 \neq 0$ the same $m^2_\beta=0.0042$ can be used (see \fig{bvk}-a).
While $\varphi_n$ needs a bit more complicated parametrization
\beq \label{EF7}
  \varphi_n\,\,=\,\,a_{\varphi,0} \,+\, a_{\varphi,1}\,n \,+\, a_{\varphi,2} \Lb n - n_\varphi\Rb^3
\eeq
For $\kmax = 10^{65}$ we obtain the following values for parameters $a_{\varphi,i}$:
\beq \label{PAM}
   a_{\varphi,0} = 0.486\;(1.520);~~
   a_{\varphi,1} = 0.0350\;(-0.0223);~~
   a_{\varphi,2} = 0.425\times10^{-4}\;(1.211\times10^{-4});~~
   n_\varphi     = 21.71\;(21.76)
\eeq
In \eq{PAM} the values of $a_{\varphi,i}$ for the case $m \neq 0, m_0\neq0$ are given in parentheses.

\begin{figure}[ht]
\begin{tabular}{cc}
  \includegraphics[width=0.5\textwidth]{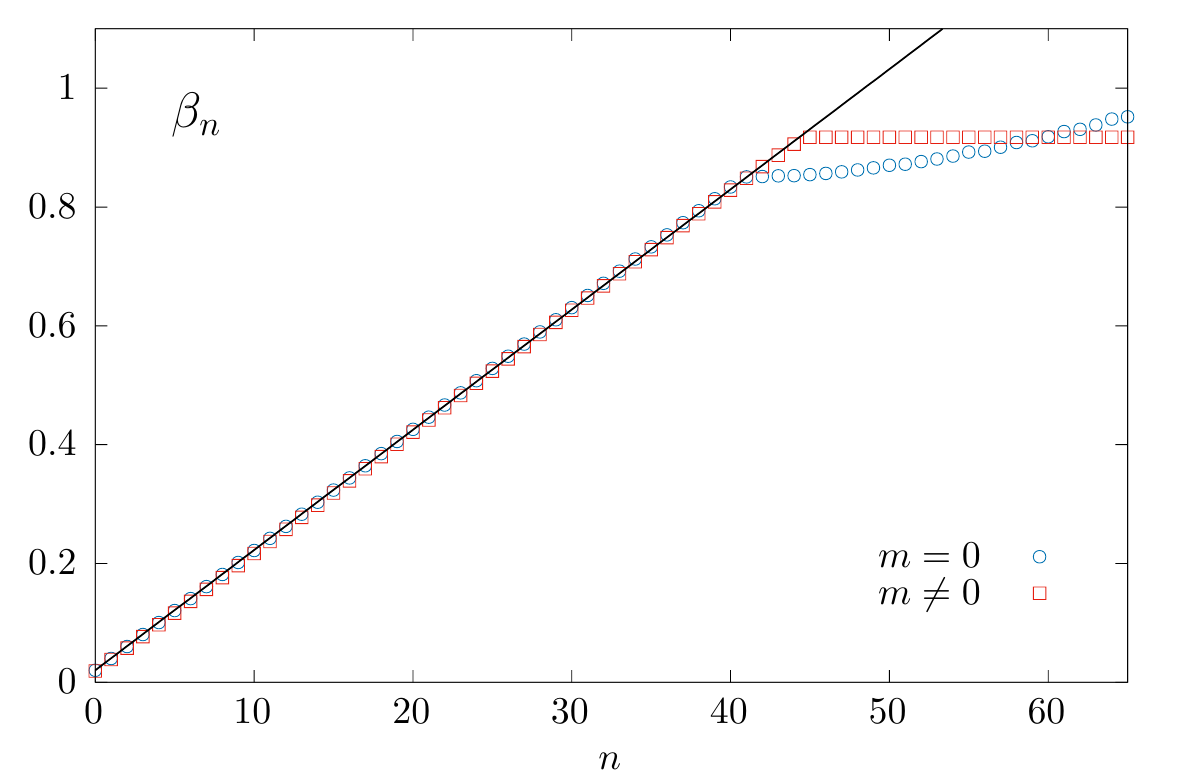} &
  \includegraphics[width=0.5\textwidth]{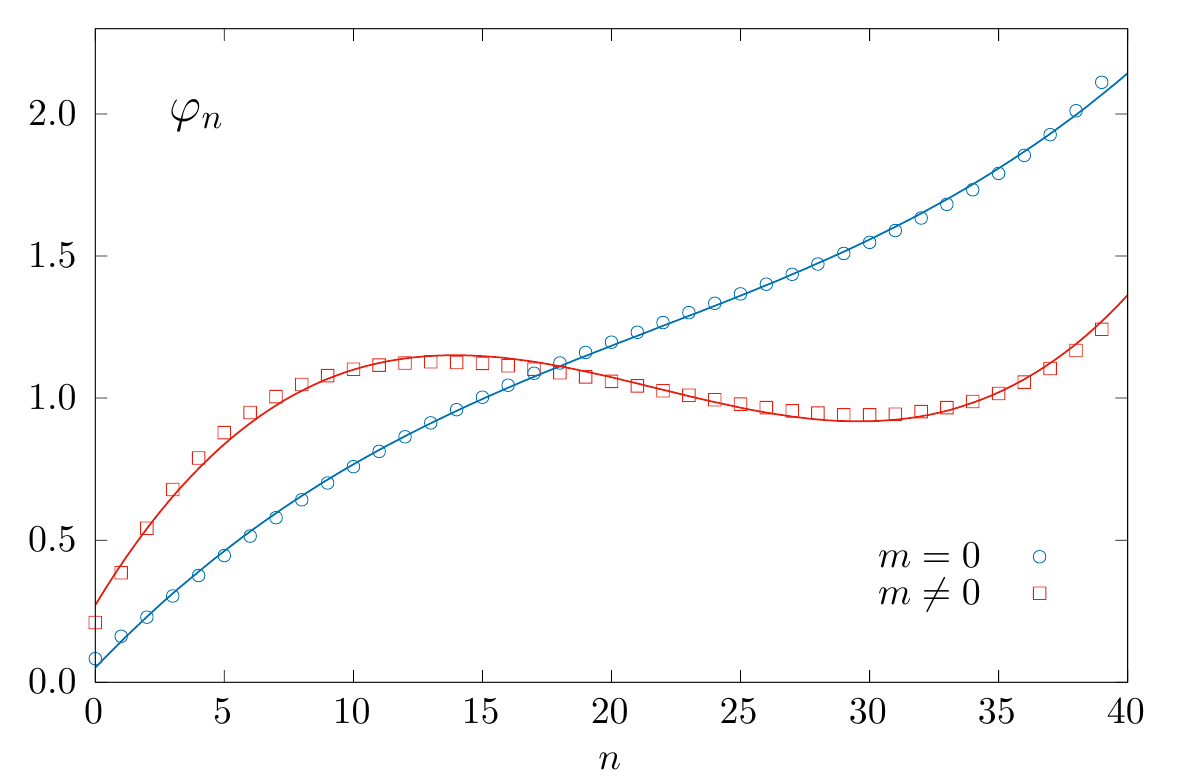}  \\
  \fig{efpar}-a &\fig{efpar}-b\\
\end{tabular}
\caption{
  Parameters $\beta_n$ (\fig{efpar}-a) and $\varphi_n$ (\fig{efpar}-b) of \eq{EF4} for the
  eigenfunctions $\phi_n(\kappa)$ versus $n$. Presented data correspond to the eigenfunctions
  obtained with $\kmin=10^{-10}$ and $\kmax=10^{65}$. One can see that $\beta_n$ has simple linear
  dependence while $\varphi_n$ has a more complicated form given by \eq{EF7}. Solid lines correspond
  to the proper parameterizations (i.e. \eq{EF6} for $\beta_n$ and \eq{EF7} for $\varphi_n$).
}
\label{efpar}
\end{figure}

\fig{efpar} shows the $n$ dependence of $\beta_n$ and $\phi_n$ for $n \leq 40$. One can see that the
linear dependence of \eq{EF6} holds for $\beta_n$, but $\varphi_n$ shows a more complicated pattern:
$\varphi_n \,\propto\,n$ with variations described by \eq{EF7}.

\fig{bvk}-a shows the dependence of $a_\beta$ on the value of $\kappa_{\rm max}$ for Gribov's
($m=0$) and lattice QCD ($m\neq 0$) gluon propagators. One can see that \eq{EF6} describes
this dependence quite well.

As far as dependence on $\kmax$ of $\beta_n $, $\varphi_n$ and $a_n$ is concerned, we found
that they have quite a simple scaling property, which allows to relate their values obtained
with different $\kmax$: values of some parameter $P_n=P(n;{\kmax}_1)$ corresponding to
${\kmax}_1$ fit well to results $P(n;{\kmax}_2)$ after simple change of the scale of $n$
\beq
  \label{Skmax}
  P(n;{\kmax}_1) \approx P(n S;{\kmax}_2), ~~~~ S = \frac{\ln({\kmax}_2)}{\ln({\kmax}_1)}
\eeq

\begin{figure}[ht]
\begin{tabular}{cc}
  \includegraphics[width=0.5\textwidth]{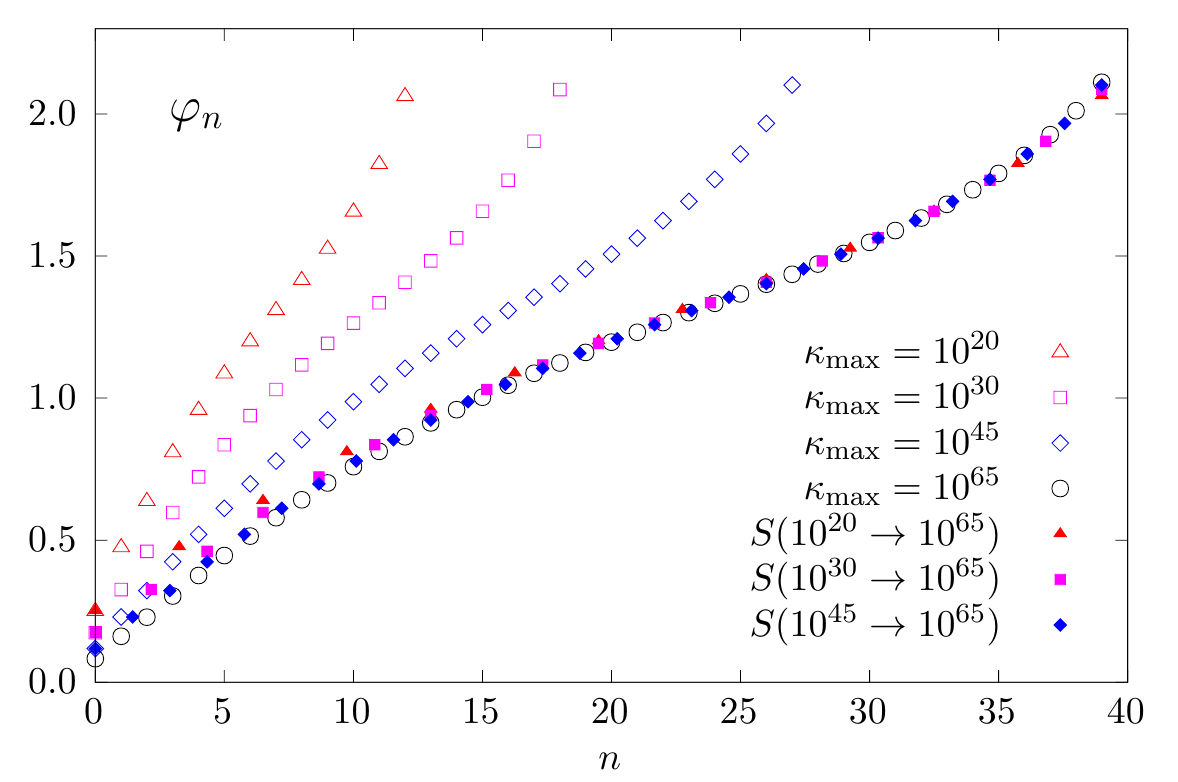} &
  \includegraphics[width=0.5\textwidth]{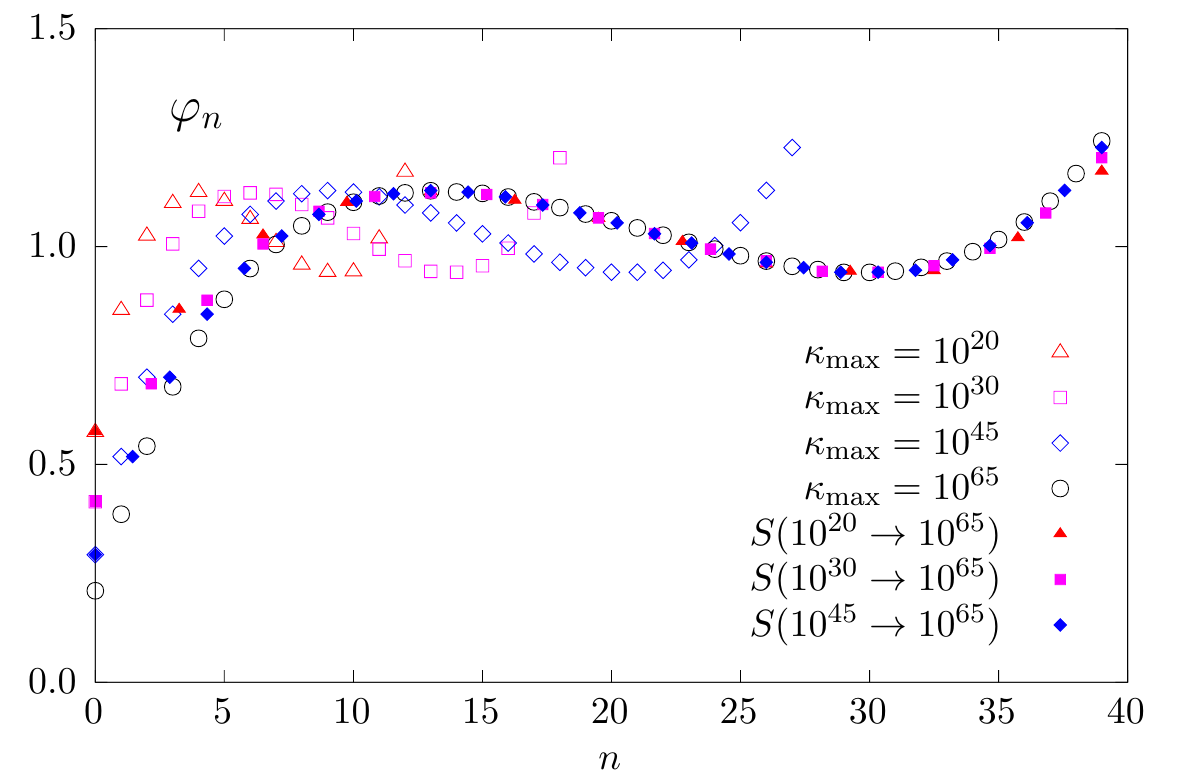} \\
  \fig{Sphi}-a &\fig{Sphi}-b\\
\end{tabular}
\caption{
  Scaling on $\kmax$ for parameter $\varphi_n$. Values on Fig.~\ref{Sphi}-a correspond
  to the Gribov's gluon propagator and on Fig.~\ref{Sphi}-b to the lattice QCD propagator.
  Open symbols denote parameter obtained at different $\kmax=10^{20}, 10^{30}, 10^{45}, 10^{65}$.
  Sets of $\varphi$ scaled on $n$ to $\kmax=10^{65}$ are denoted by appropriate solid symbols.
}
\label{Sphi}
\end{figure}

Such scaling behaviour for $\beta_n$ follows from \eq{EF6}, while \fig{Sphi} illustrates
such $\kmax$ scaling for parameter $\varphi_n$. Figure shows parameters obtained at
different $\kmax$ in the range from $10^{20}$ to $10^{65}$. Sets of $\varphi$ were
scaled on $n$ with proper coefficient (see \eq{Skmax}) to $\kmax=10^{65}$. One can see
that "scaled" points are in good agreement with original values for $\kmax=10^{65}$.

For the Gribov's propagator the eigenfunction $\phi_n\Lb \kappa\Rb \,\,\propto\,\,\kappa$
in the region of small $\kappa$. In other words, the eigenfunctions in the coordinate space
exhibit the power-like decrease at long distances.

\begin{figure}[ht]
\begin{tabular}{cc}
  \includegraphics[width=0.5\textwidth]{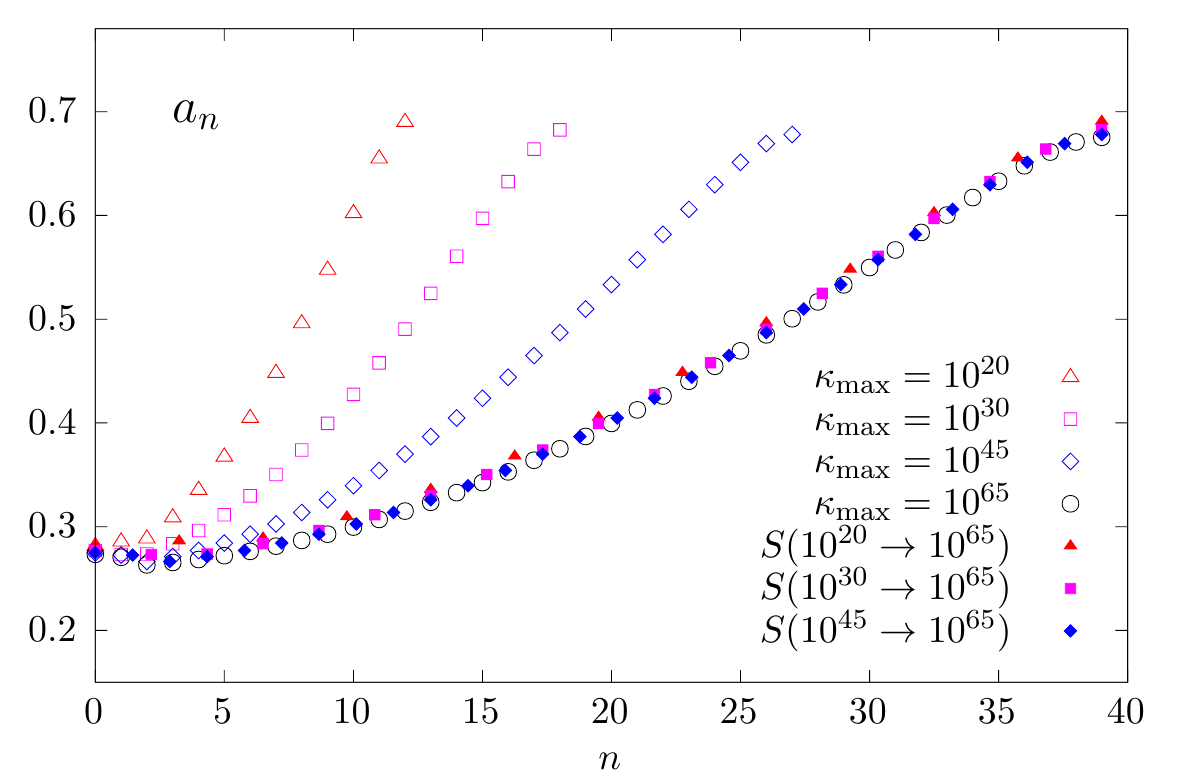} &
  \includegraphics[width=0.5\textwidth]{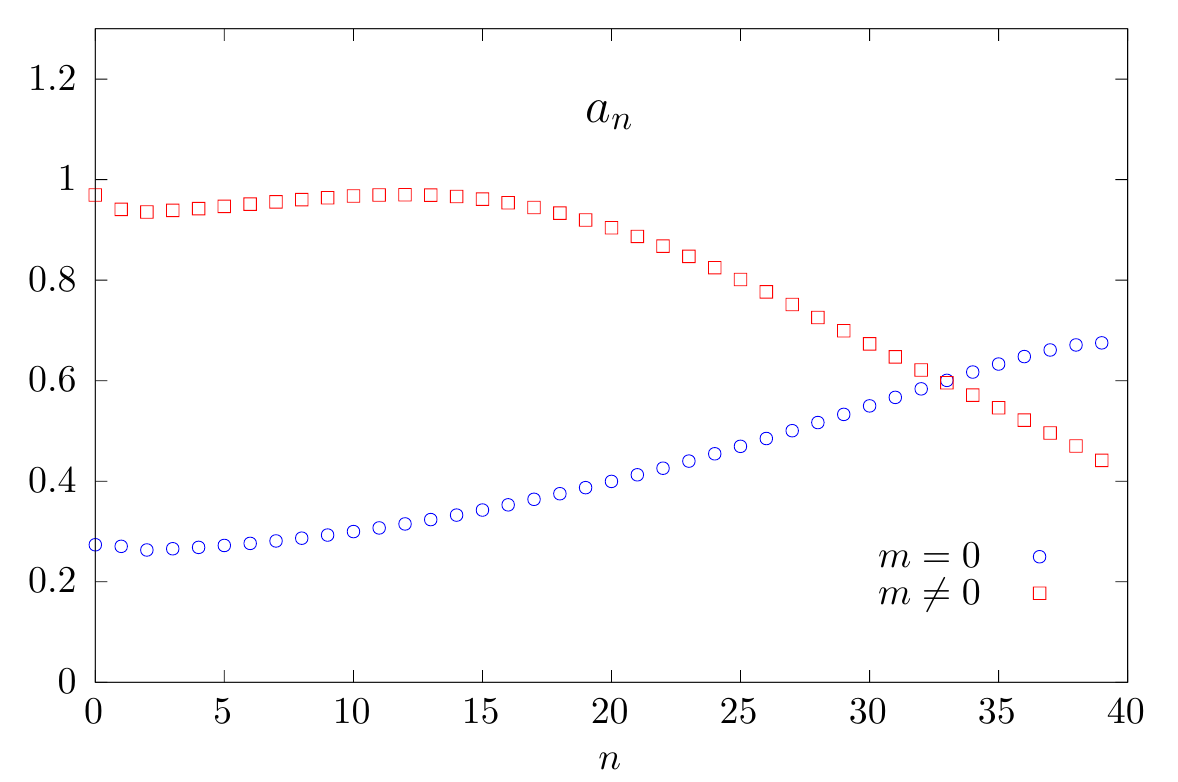} \\
  \fig{an}-a &\fig{an}-b\\
\end{tabular}
\caption{
  Values of parameter $a_n$ from the \eq{EF4} versus $n$. Left plot (\fig{an}-a) demonstrates
  scaling property of this parameter for the case with Gribos's propagator. Right plot (\fig{an}-b)
  shows $a_n$ for both variants (i.e. Gribov's and QCD propagators) corresponding to eigenfunctions
  with $\kmin=10^{-10}$ and $\kmax=10^{65}$.
}
\label{an}
\end{figure}

It turns out that for small $\kappa$ we can use \eq{EF4}, which introduce the dependence of
parameter $a_n$ on $n$. \fig{an}-a shows the scaling behaviour of \eq{Skmax} for  $a_n$ in
the case of Gribov's propagator. It should be stressed that the same scaling behaviour holds
for the case of $ m \neq 0$.

\fig{an}-b shows the dependence of $a_n$ on $n$. From this figure we see that $a_n \,\propto\,n^2$
for the Gribov's propagator with $a_n = 0.3$ at $n  = 0$. In other words, the typical $\kappa$
turns out to be in the region of $\kappa = 0.27\div\,0.75$. It should be stressed that \eq{EF4}
describes quite well the behaviour of the eigenfunctions both at large $\kappa \,\geq\,1$ and
at small $\kappa \,\leq\,a_n$, but at $\kappa \,\sim\,a_n$ \eq{EF4} does not lead to a good fit
of the eigenfunctions. For the case of the lattice QCD propagator (\eq{GZ3} with $m =1.27$ and
$m_0$=3.76) \fig{an}-b shows that $a_n$ decreases with $n$ approaching $a_n \approx 0.4 $ at
$n \to 40$. One can see that for small $n$ the typical values of $\kappa \sim\,1$. and the range
of typical $\kappa$ is $0.4 \div 1$. We would like to stress that the value of $a_n$ cannot be
viewed as a typical scale for the $\kappa$ dependence of the eigenfunctions. Indeed, one can see
directly from \fig{ef} that the typical $\kappa $ is about $\kappa \sim 1$ for both cases.

\subsection{Eigenfunctions for $E_n\,\,=\,E_0\,\,=\,T\Lb \kappa = 0\Rb$}

As we have discussed above, the solution leads to the multiple degenerate state at
$E_n = E_0 = T\Lb \kappa = 0\Rb$. Using \eq{EVN0} we can estimate the value of $\beta^0$: viz.
$E_0 = - 2 \,\psi(1) \,+\,\psi\Lb \h \,+\,i\,\beta^0\Rb \,+\,\psi\Lb \h \,-\,i\,\beta^0\Rb$.
Corresponding eigenfunction index $n$, where the first degenerate eigenvalue appears,
can be estimated using \eq{EF6}: degenerate eigenfunction sequence starts when $\beta_n$
reaches value $\beta^0$. For the Gribov's propagator $\beta^0\approx0.85$ (see \fig{efpar}-a)
leads to the value $n=41$ in the case with $\kmax=10^{65}$, while for RGZ gluon propagator
$\beta^0\approx0.92$ gives $n=45$ (see \fig{efpar}-a). At such values of $n$ the $\kappa$
behaviour of the eigenfunctions shows a discontinuity in $\kappa$: of course, numerical
values of eigenfunctions at $\ln\Lb\kappa/\kappa_{\rm min}\Rb = n \Delta$ (see \eq{NS2})
are still finite, but the values in the neighbouring nodes have a different sign and
derivative, indicating that eigenfunctions have pole in $\kappa$ located somewhere
between nodes.

\begin{figure}[ht]
\begin{tabular}{cc}
  \includegraphics[width=0.5\textwidth]{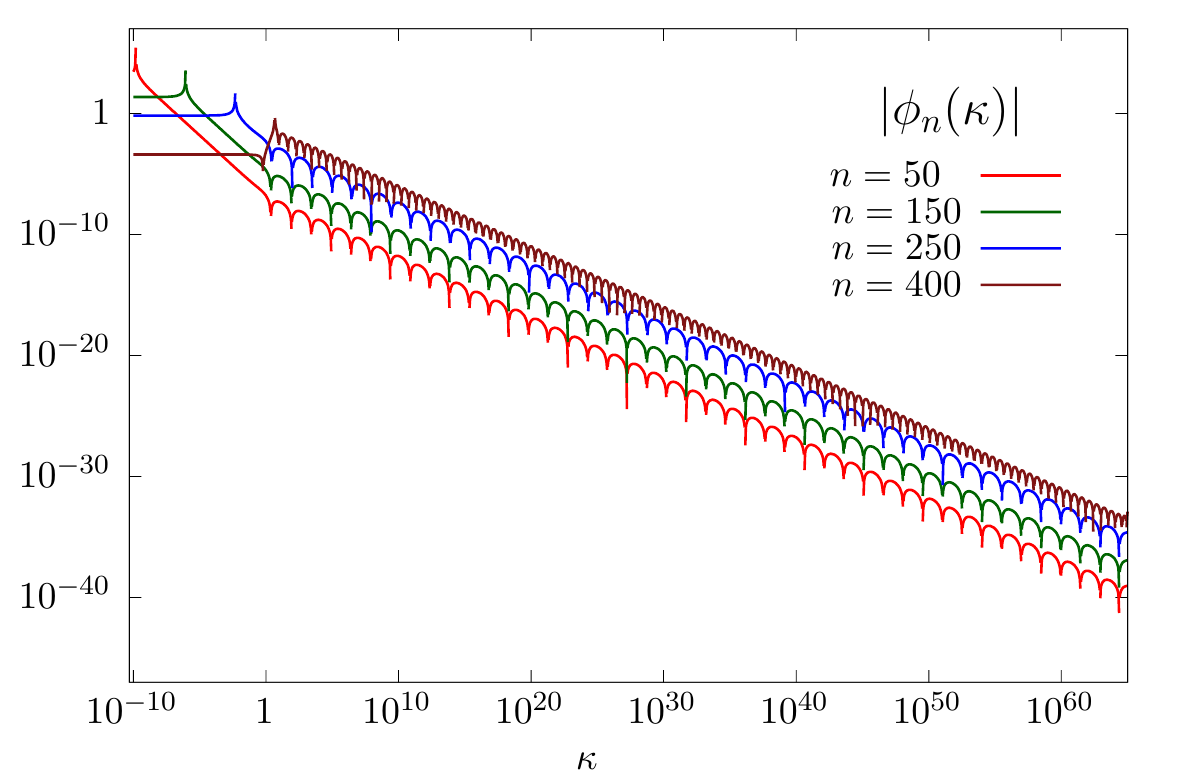} &
  \includegraphics[width=0.5\textwidth]{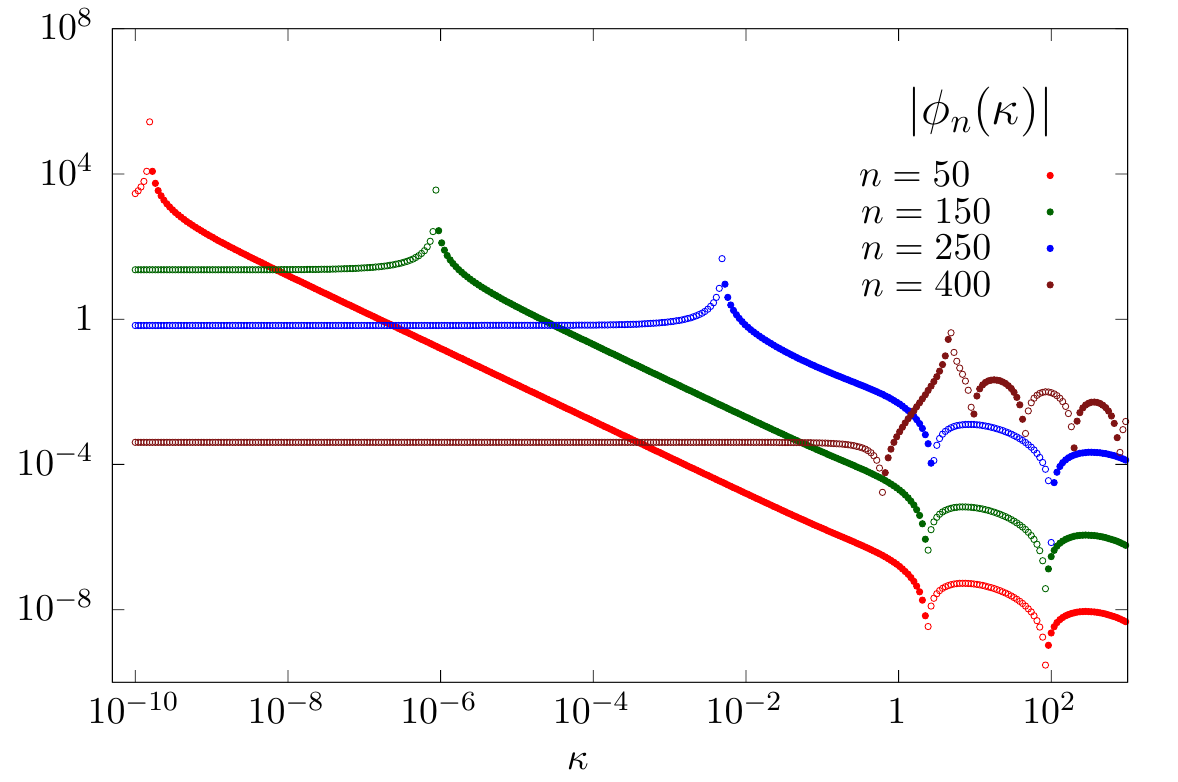} \\
  \fig{efe0mass}-a &\fig{efe0mass}-b\\
\end{tabular}
\caption{
  The examples of the eigenfunctions with the eigenvalue $E_n = E_0 = T\Lb \kappa=0\Rb$ for
  the lattice QCD gluon propagator(\eq{GZ3} with $m=1.27$ and $m_0=3.76$). \fig{efe0mass}-a
  shows $\phi_n\Lb \kappa\Rb$ versus $\kappa$ with $n=50,150,250$ for which $E_n = E_0$ and
  $\phi_{400}\Lb \kappa\Rb$ with $E_{400} > E_0$. One can see that all eigenfunctions with
  $E_n = E_0$ have the same number of zeros. The same eigenfunctions are shown in \fig{efe0mass}-b
  but in the region of small $\kappa \leq 100$. It is clearly seen that $\phi_n\Lb \kappa\Rb$
  have the pole whose position moves from $\kappa_{\rm min}$ to $\kappa \approx 1$. Function
  $\phi_{400}\Lb \kappa\Rb$ has a different number of zeroes, which corresponds to increase of
  the value of $\beta$ in \eq{EFE01}.
}
\label{efe0mass}
\end{figure}

The structure of the eigenfunctions with this eigenvalue is rather simple for the
lattice (RGZ) QCD gluon propagator (see \eq{GZ3} with $m =1.27$ and $m_0 = 3.76$) and
it is close to one, that has been discussed in Ref.\cite{LLS} for \eq{BFKLHIGGS}.
The eigenfunctions for $E_n=E_0$ have poles in $\kappa=\kappa_{p,n}$ as it has been shown
in section III-C. Actually the minimal value of $\kappa_{p,n}$ is equal to $\kmin$ (strictly
speaking the first pole is located somewhere between the first node $\kappa_0=\kmin$ and
the next node $\kappa_1 = \kmin\exp(\Delta_\kappa) \approx \kmin (1+\Delta_\kappa$). With
each increase of $n$ pole moved exactly to the next interval on $\kappa$ (i.e. second pole
is located between $\kappa_1$ and $\kappa_2$ and so on). This sequence terminates when the pole
reaches the maximal value of $\kappa=\kappa^0\,\approx\,3$, where $\kappa^0$ is the location
of the first zero of the eigenfunction of \eq{EF4}. All eigenfunctions with $E_n = E_0$ have
the same number of zeros. All these features can be seen from \fig{efe0mass}.

\begin{figure}[ht]
\begin{tabular}{cc}
  \includegraphics[width=0.5\textwidth]{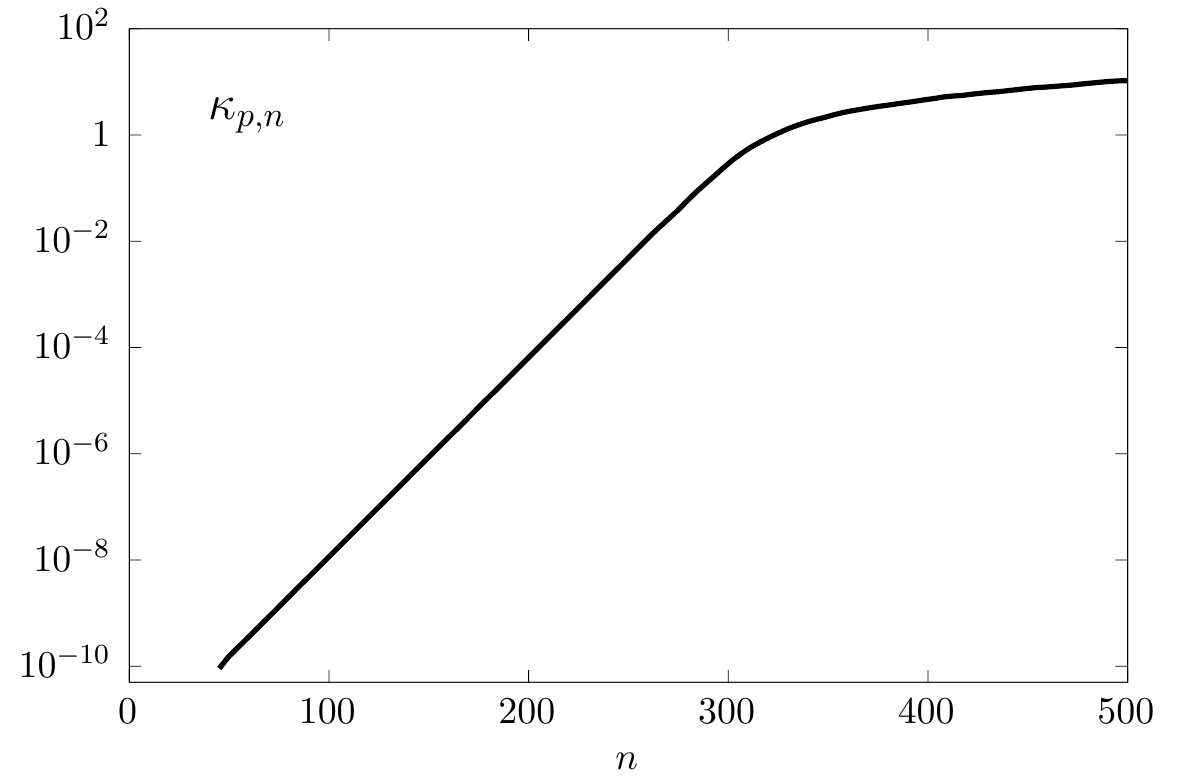} &
  \includegraphics[width=0.5\textwidth]{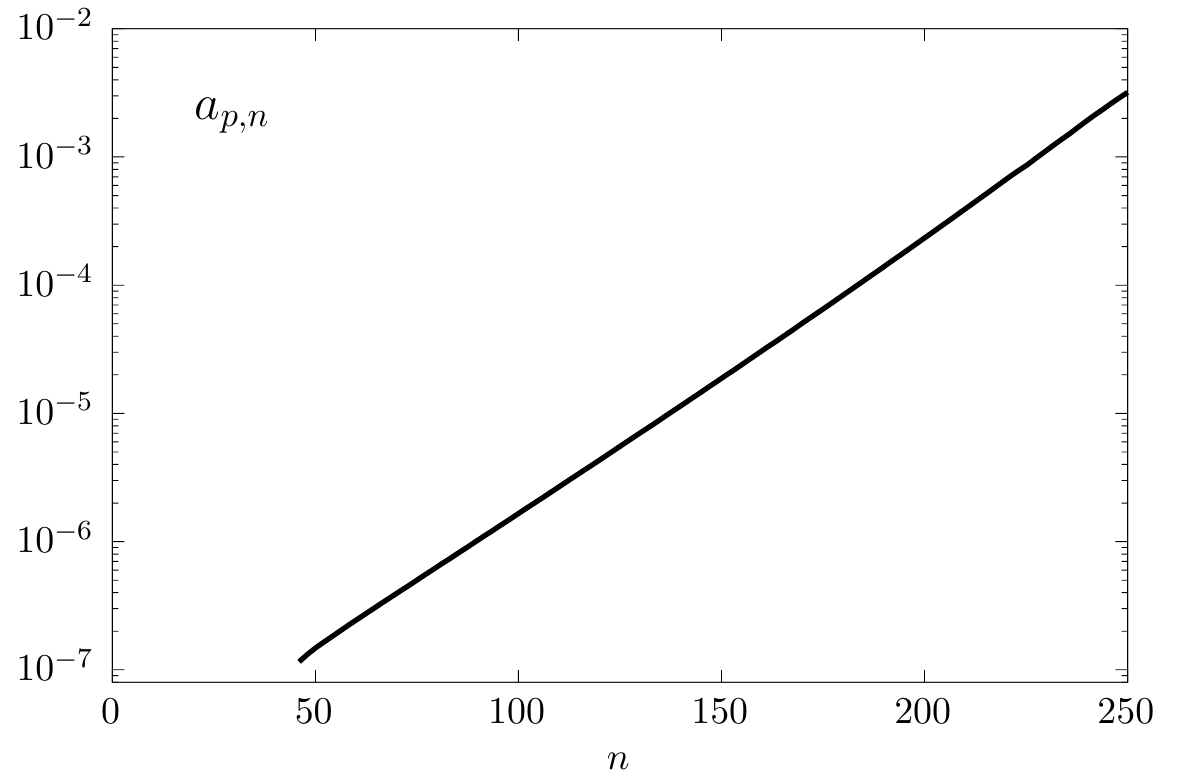} \\
  \fig{parwfmass}-a &\fig{parwfmass}-b\\
    \includegraphics[width=0.5\textwidth]{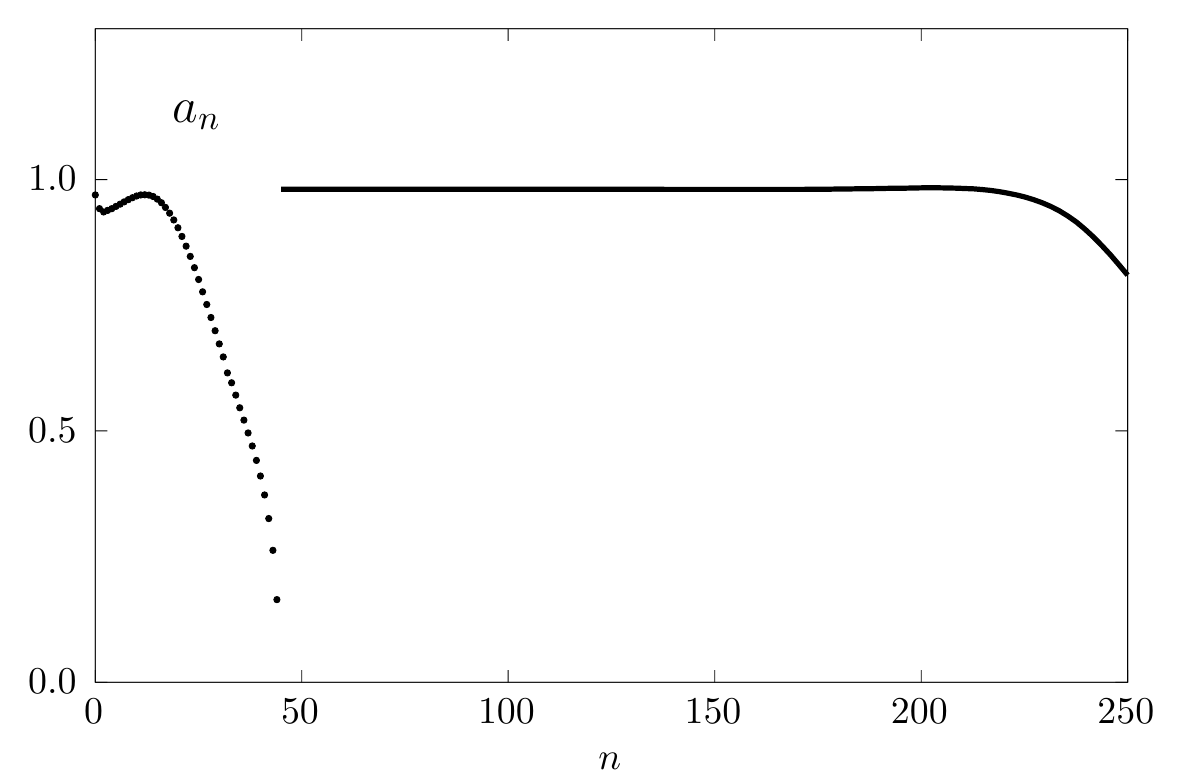} &
    \includegraphics[width=0.5\textwidth]{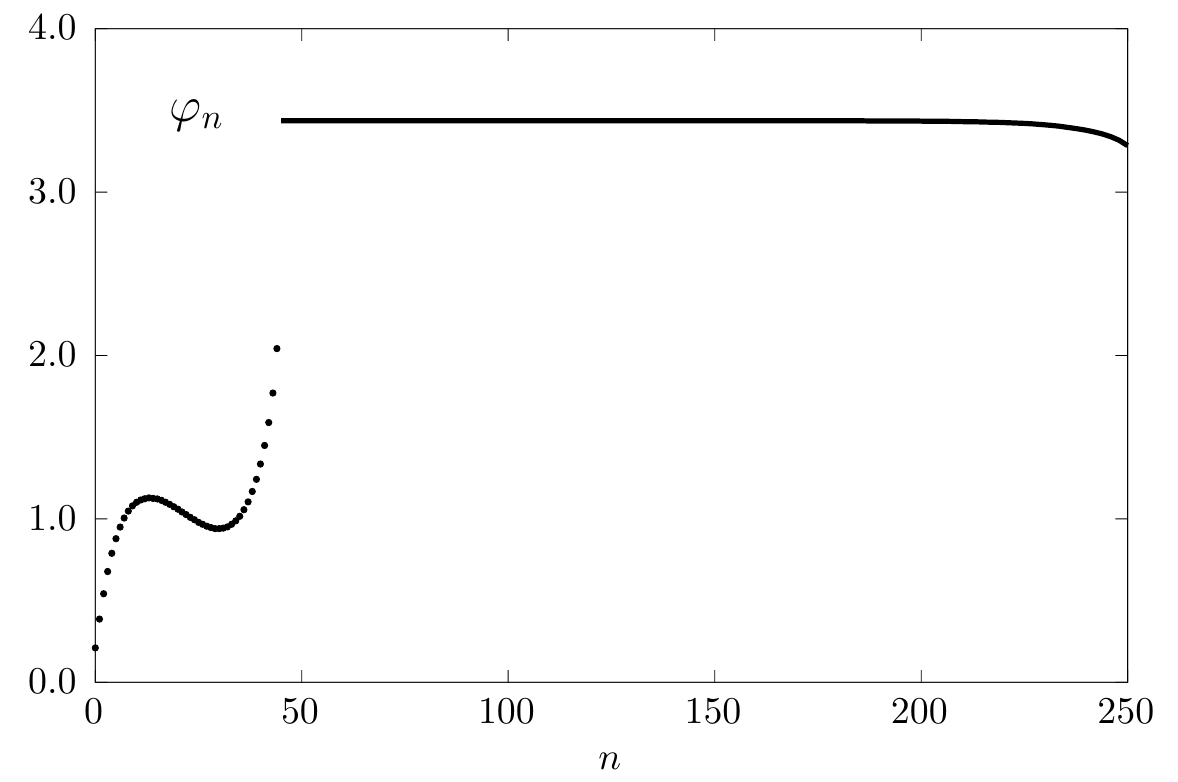} \\
  \fig{parwfmass}-c &\fig{parwfmass}-d\\
\end{tabular}
\caption{
  Parameters of the wave functions with $E_n = E_0$ versus $n$. \fig{parwfmass}-a shows
  the position of the pole $k_{p,n}$ in \eq{EF4} as a function of $n$ while \fig{parwfmass}-b
  presents the $n$ dependence of the residue $a_{p,n}$. \fig{parwfmass}-c and \fig{parwfmass}-d
  describe the dependence of parameters $a_n$ and $\varphi_n$ on $n$.
}
\label{parwfmass}
\end{figure}

Generally for $m \,\neq\,0$ and $m_0\,\neq\,0$, the eigenfunction can be approximated
by the following expression:
\beq \label{EFE01}
\phi^{\mbox{\tiny (approx)}}_n\Lb\kappa\Rb
  \,=\,\frac{a_{p,n}}{\kappa_{p,n}\,-\,\kappa} \,+\,\phi_n\Lb\kappa;\,\eq{EF4}\Rb
  \,=\,\frac{a_{p,n}}{\kappa_{p,n}\,-\,\kappa} \,+\,\frac{\alpha_n\,(\kappa+m)}{\sqrt{(\kappa\,+\,a_n)^3}}
     \,\sin\Lb\beta^0 Ln(\kappa)\,+\,\varphi_n\Rb
\eeq
\eq{EFE01}, having $\beta^0$ (see \fig{bvk}-b), which does not depend on $n$, reflects the fact
that for all these states the behaviour of the eigenfunctions at large $\kappa \,\geq\,1$ can
be described by one function with the same number of zeros. \fig{parwfmass} shows the $n$-dependence
of other parameters of $\phi^{\mbox{\tiny (approx)}}_n\Lb \kappa\Rb$ (see \eq{EFE01}). One can see
that both the position of the pole $\kappa_{p,n}$ and its residue $a_{p,n}$ are proportional to $n$
($\ln\Lb\kappa_{p,n}\Rb \,\propto\,n, \ln\Lb a_{p,n}\Rb\, \propto\, n$), while the parameters
of $\phi_n\Lb \kappa; \eq{EF4}\Rb $: $a_n$ and $\varphi_n$ do not depend on $n$ in the range
$n = 50 - 250$ which corresponds to $E_n = E_0$. The value of $a_n=1$ gives us the typical
transverse momentum $q\,=\,\mu$ (see \eq{VAR}).

\begin{figure}[ht]
\begin{tabular}{cc}
  \includegraphics[width=0.5\textwidth]{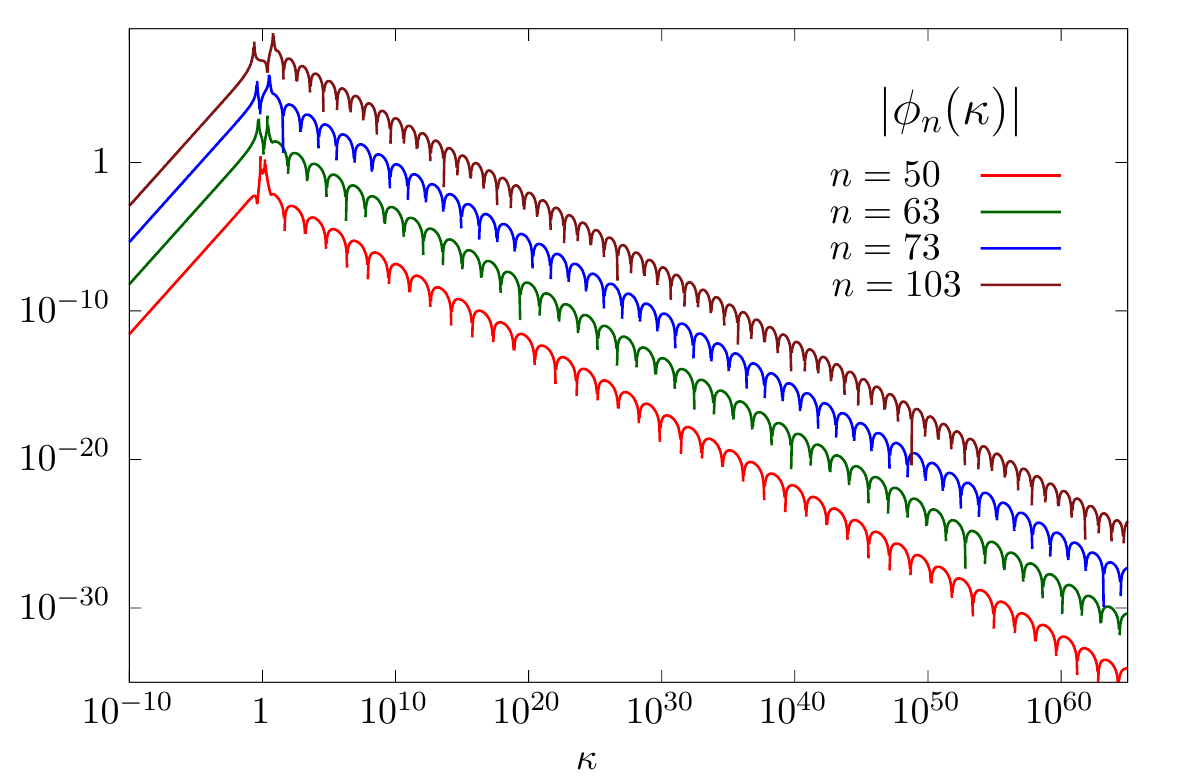} &
  \includegraphics[width=0.5\textwidth]{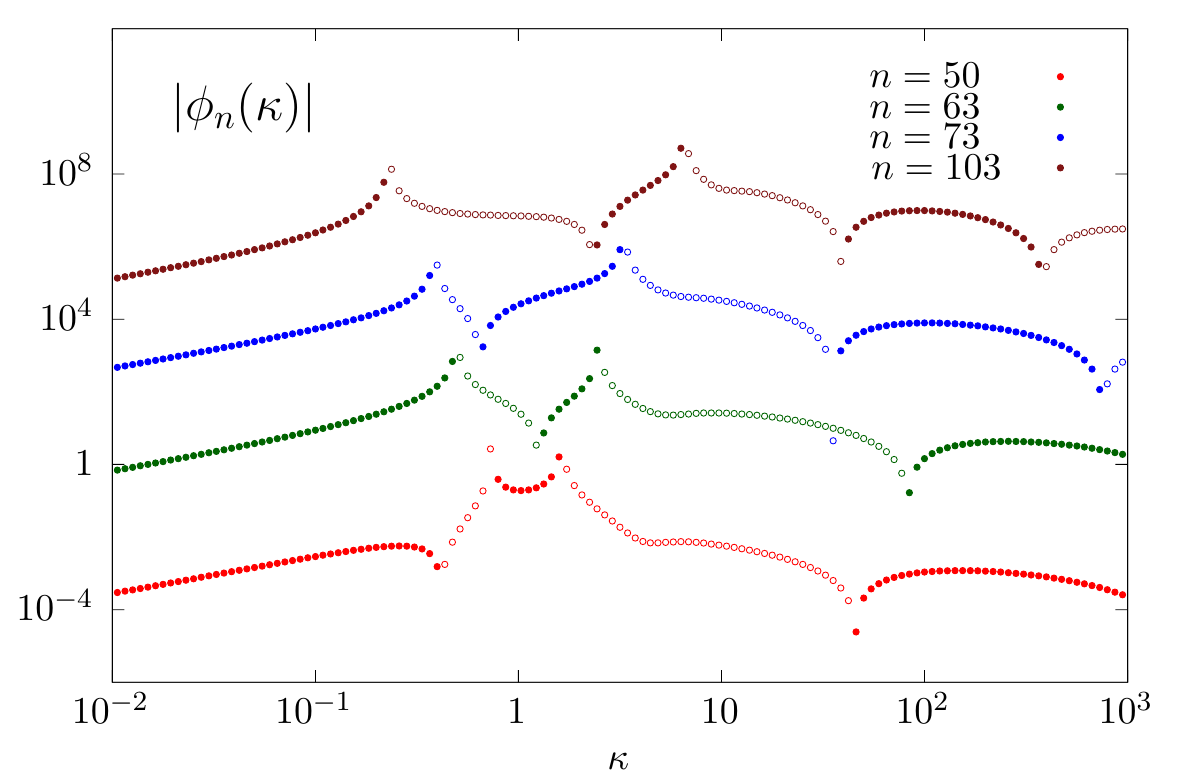} \\
  \fig{efe0}-a &\fig{efe0}-b\\
\end{tabular}
\caption{
  The examples of the eigenfunctions with the eigenvalue $E_n = E_0 = T\Lb \kappa=0\Rb$
  for Gribov's gluon propagator(\eq{GZ3} with $m=0$ and $m_0=0$). \fig{efe0}-a shows
  $\phi_n\Lb \kappa\Rb$ versus $\kappa$ with $n=50,63,73$ for which $E_n \approx E_0$
  and $\phi_{103}\Lb\kappa\Rb$ with $E_{103} > E_0$. (For better clarity, some functions
  were scaled: values for $n=63$ are multiplied by $10^3$, $n=73$ by $10^6$ and $n=103$
  by $10^9$.) One can see that all eigenfunctions with $E_n = E_0$ have approximately
  the same number of zeros, but $\beta$ in \eq{EFE02} is not a constant but slowly grows
  as $n$ increase. In \fig{efe0}-b the same functions are shown in the region of small
  $\kappa \leq 100$. It is clearly seen that $\phi_n\Lb \kappa\Rb$ have two poles whose
  position moves in opposite directions from $\kappa=1$. The open circles in \fig{efe0}-b
  show the region where the function is negative.
}
\label{efe0}
\end{figure}

For Gribov's gluon propagator the structure of the eigenfunction with $E_n=E_0$ is
much more complex. First, one can notice from \fig{efe0} that the number of zeros
are not the same for these eigenfunction but $\beta^0$ in \eq{EFE02} changes with
$n$ rather slowly (see \fig{efpar}-a for $n$ in the region $n = 41 \div 70$).

Second, we see that $\phi_n\Lb \kappa\Rb$ with $E_n = E_0$ have two poles. First pair
of poles $\kappa_{p,1}<\kappa_{p,2}$ appear near $\kappa\approx1$. With increase of $n$
the smaller one ($\kappa_{p,1}$ decreases, while $\kappa_{p,2}$ increases. On each
increment of $n$ only one of $\kappa_{p,i}$ moves to the neighbouring $\kappa$ interval
between nodes. So the distance between these poles (in terms of index of $\kappa$-nodes)
each time increases exactly on $1$. The contribution of each of these poles vanishes at
$\kappa \to 0$ and residues of these poles can have the same or opposite sign.

Third, the position of the poles are in the region $\kappa = 0.1 \div 10$ and they exist
also in the eigenfunctions with $E_n > E_0$. Fourth, two poles in the eigenfunctions have
close positions. \eq{EFE02} reflects the main features that we have discussed but
the actual structure of the eigenfunction turns out to be much more complex.

However, for the Gribov's propagator ($m = 0, m_0 =0$) the eigenfunction vanishes at $\kappa=0$
and can be approximated as follows:
\beq \label{EFE02}
  \phi^{\mbox{\tiny (approx)}}_n\Lb \kappa\Rb\,\,=
       \,\,\frac{\kappa\,a_{p,1}(n)}{\kappa^2 - \kappa^2_{p,1}(n)}\,\,
    \pm\,\,\frac{\kappa\,a_{p,2}(n)}{\kappa^2 - \kappa^2_{p,2}(n)}\,\,
     + \,\,\frac{\alpha_n\,\kappa}{\sqrt{(\kappa+a_n)^3}}\sin\Lb\beta^0 \, Ln\kappa\,+\,\varphi_n\Rb
\eeq

\begin{figure}[ht]
\begin{tabular}{cc}
  \includegraphics[width=0.5\textwidth]{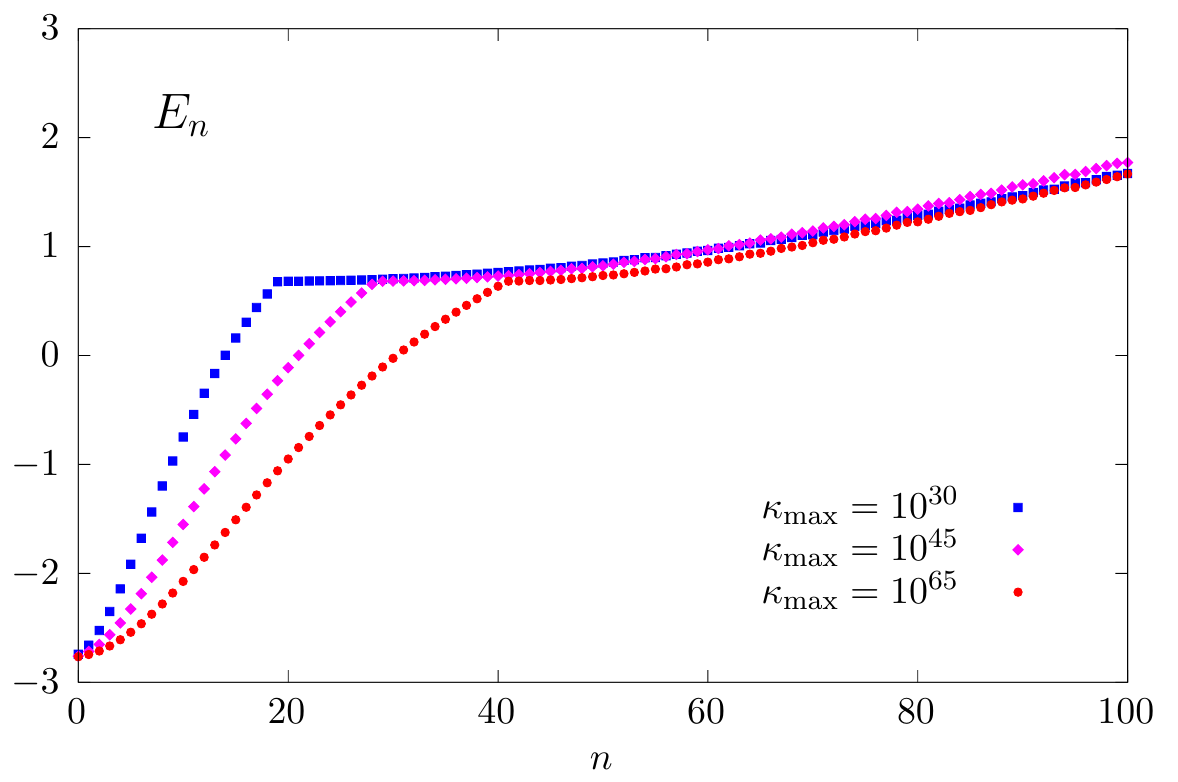} &
  \includegraphics[width=0.5\textwidth]{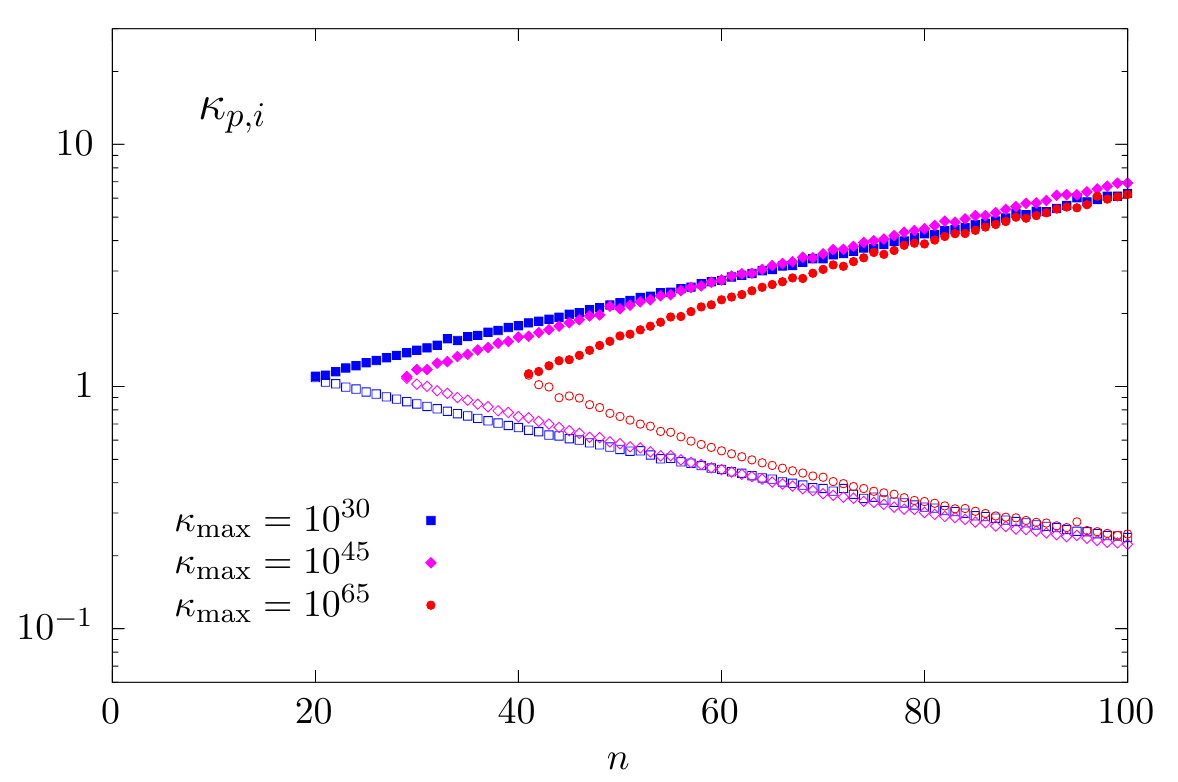} \\
  \fig{efgre0}-a &\fig{efgre0}-b\\
\end{tabular}
\caption{
  The eigenvalues for the Gribov's gluon propagator (see \eq{GZ3} with $m=0$ and $m_0=0$)
  versus $n$ (\fig{efgre0}-a). The value of $n$ at which energy $E_n$ reaches $E_0$,
  can be found from the equation $\beta_n = \beta^0$= 0.85. \fig{efgre0}-b shows
  the dependence of two poles in \eq{EFE02} on $n$ at different values of $\kmax$.
  Note, that the minimal value of $n$ when two poles $\,\kappa_{p,i}$ appear is
  determined by the same condition $\beta_n = \beta^0$= 0.85.}
\label{efgre0}
\end{figure}

The appearance of two poles in \eq{EFE02} looks natural (see section III-C) due to
multiple degeneracy of this eigenvalue. Indeed, due to this the  sum of two functions
with one pole in each, is also the eigenfunction. \fig{efgre0} demonstrates that
the region of $n$ for the degenerate states with $E_n = E_0$ is very narrow but
the structure of the eigenfunction with two poles lasts for $E_n \,>\,E_0$.

 \subsection{Eigenfunctions for $E_n\,\,>\,\,E_0\,\,=\,T\Lb \kappa = 0\Rb$}

For $E_n \,\,>\,\,E_0$ the eigenfunctions take the following form for $\kappa\,>\,1$:
\beq \label{EFE03}
  \phi_n\Lb \kappa\Rb\,\,=\,\,\frac{\alpha_n}{\sqrt{\kappa}}\,\sin\Lb \beta_n Ln(\kappa) \,+\, \varphi_n\Rb
\eeq
with $\beta_n \,\propto\,n$ for both Gribov's and lattice QCD gluon propagators as it is
seen from \fig{bvk}-b. One can see from \fig{rootgen}-a, that for$\kappa\,\gg\,1$ $E_n$
tends to the same asymptotic values for all four equations that we have studied in this
paper. For $\kappa\,<\,1$ the eigenfunctions have the similar structure as for $E_n = E_0$,
i.e. they have one pole (see \fig{parwfmass}-a) for the lattice gluon propagator and two
poles for the Gribov's one (see \fig{efgre0}-b). Since these eigenvalues corresponds to
the Pomeron intercept $\omega_\pom \,=\, -\bas E_n\,\,<\,\,\omega_0=- \bas \,E_0 \,<\,0$,
they do not contribute to the high energy behaviour of the scattering amplitude.

\section{The scattering amplitude}
\subsection{Green's function of the BFKL Pomeron for the Gribov-Zwanziger confinement}

The Green's function of the BFKL equation on $Y=\,\ln(1/x)$ representation takes the general form
\beq \label{GF1}
  G\Lb Y, \kappa_{\mbox{\tiny fin}} | 0, \kappa_{\mbox{\tiny in}}\Rb
  \,\,=\,\,
    \sum_{n=0}^{\infty}\int^{\epsilon\,+\,i\,\infty}_{\epsilon \,-\,i\,\infty}
    \frac{d \omega}{ 2\,\pi\,i}\,\,
    \frac{1}{\omega\,-\,\omega_n}\,\,
    e^{\omega\,Y}\,
    \phi_n\Lb \kappa_{\mbox{\tiny in}}\Rb\,
    \phi_n\Lb \kappa_{\mbox{\tiny fin}}\Rb
  \,\,=\,\,
    \sum_{n=0}^{\infty}\,e^{-\bas E_n\,Y}\,
    \phi_n\Lb \kappa_{\mbox{\tiny in}}\Rb\,
    \phi_n\Lb \kappa_{\mbox{\tiny fin}}\Rb
\eeq
At high energies the main contribution stems from the minimal energy. However, we cannot
restrict ourselves by calculating only one term in \eq{GF1}. To demonstrate this, we use
\eq{EF4} for the approximate eigenfunctions, which can be written in the following form:
\beq \label{GF2}
  \phi_n^{\mbox{\tiny approx}} \,\,=\,\,\Phi_n\Lb \kappa\Rb \,\sin\Lb a_\beta\,(n\,+1)\, Ln(\kappa)\,+\,\varphi_n\Rb
\eeq
where
\beq \label{GF3}
  \Phi_n\Lb \kappa\Rb\,\,\,=\,\,\,\frac{\alpha_n\,(\kappa+m)}{\sqrt{(\kappa\,+\,a_n)^3}}
\eeq

The eigenvalues of \eq{EVN0} we calculate in diffusion approximation in which:
\beq  \label{EVND}
\omega\Lb n\Rb
  \,\,=\,\,-\bas\,E_n
  \,\,=\,\,\omega_{\mbox{\tiny BFKL}}\,\,-\,\,D\,a_\beta^2\,n^2\,\,+\,\,{\cal O}\Lb n^3\Rb
  \,\,=\,\,\omega_{\mbox{\tiny BFKL}}\,\,-\,\,D\,\beta^2
\eeq
where $\omega_{\mbox{\tiny BFKL}}\,=\,4\,\ln2\, \bas$; $ D\,=\,14\, \zeta(3) \,\bas$.
Therefore, in this approximation the Green function takes the form
\beq \label{GF4}
  G\Lb Y, \kappa_{\mbox{\tiny fin}} | 0, \kappa_{\mbox{\tiny in}}\Rb
    \,\,=\,\,e^{\omega_{\mbox{\tiny BFKL}}\,Y}
         \,\sum_{n=0}^{\infty}
           \phi_n\Lb\kappa_{\mbox{\tiny fin}}\Rb\,
           \phi_n\Lb\kappa_{\mbox{\tiny  in}}\Rb
         \,e^{-D\,Y\,a_\beta^2\,n^2}
    \to\,\,e^{\omega_{\mbox{\tiny BFKL}}\,Y}\,\int^{\infty}_0 \,d\beta
        \,\phi\Lb\kappa_{\mbox{\tiny fin}},\beta\Rb\,
          \phi\Lb\kappa_{\mbox{\tiny  in}},\beta\Rb
        \,e^{ - D\,Y\,\beta^2}
\eeq
Taking the integral over $\beta$ in \eq{GF4} we obtain the following Green's function at large values of $Y$:
\beq \label{GF5}
  G\Lb Y, \kappa_{\mbox{\tiny fin}} | 0, \kappa_{\mbox{\tiny in}}\Rb
    \,\,=\Phi_0(\kappa_{\mbox{\tiny fin}})
         \Phi_0(\kappa_{\mbox{\tiny  in}})
       \,\h\,e^{\omega_{\mbox{\tiny BFKL}}Y}\,\sqrt{\frac{\pi}{D\,Y}}
       \Bigg\{
             e^{-\frac{\Lb Ln\Lb \kappa_{\mbox{\tiny fin}}\Rb
                      \,-\,Ln\Lb \kappa_{\mbox{\tiny  in}}\Rb\Rb^2}{4\,D\,a_\beta^2\,Y}}
        \,-\,e^{-\frac{\Lb Ln\Lb \kappa_{\mbox{\tiny fin}}\Rb
                      \,+\,Ln\Lb \kappa_{\mbox{\tiny  in}}\Rb\,+\,2\,a_\phi\Rb^2}{4\,D\,a_\beta^2\,Y}}
       \Bigg\}
\eeq
One can see that at large $Y$, $G\Lb Y,\kappa_{\mbox{\tiny fin}} | 0, \kappa_{\mbox{\tiny in}}\Rb
\,\,\propto\,\,\Lb D \,Y\Rb^{-3/2}\,e^{\omega_{\mbox{\tiny BFKL}}\,Y}$, which should be compared with
the massless BFKL case for which $G\Lb Y, \kappa_{\mbox{\tiny fin}} | 0, \kappa_{\mbox{\tiny in}}\Rb
\,\,\propto\,\,\Lb D \,Y\Rb^{-1/2}\,e^{\omega_{\mbox{\tiny BFKL}}\,Y}$.

These estimates show that we need to sum the contributions of the eigenvalues
in the vicinity of $n=0$. The source of such contribution can be seen from
the first two components of the sum over $n$ in \eq{GF1}, which can be
re-written as follows:
\beq
   G\Lb Y, \kappa_{\mbox{\tiny fin}} | 0, \kappa_{\mbox{\tiny in}}\Rb \,\,\propto\,\,
    \Phi_0(\kappa_{\mbox{\tiny fin}})
    \Phi_0(\kappa_{\mbox{\tiny  in}})
    e^{\omega_{\mbox{\tiny BFKL}}\,Y}\Bigg(
        1\,+\,\frac{\Phi_1(\kappa_{\mbox{\tiny fin}})\,\Phi_1(\kappa_{\mbox{\tiny in}})}{%
                    \Phi_0(\kappa_{\mbox{\tiny fin}})\,\Phi_0(\kappa_{\mbox{\tiny in}})}
            \,e^{\Delta \omega_1 \,Y}
    \Bigg)
\eeq
where $\Delta\omega_1\,=\,-\bas\Lb E_0 - E_1\Rb$.

For $\kappa_{\rm max}\to\infty$ ~ $\Delta\omega_1\to0$, however, the product $\Delta\omega_1\,Y$
at large $\kmax$ and $Y$ is undefined. Hence, we have to perform numerical estimates for the sum
of \eq{GF1} to determine the answer. We will discuss such kind of estimate below. At the moment
we wish to emphasize that since in the vicinity of $n=0$ the spectrum of the master equation
coincide with the QCD BFKL equation, one can see that the influence on the asymptotic behaviour
of the scattering amplitude due to Gribov-Zwanziger confinement is rather small. Indeed, as we
have seen from the above estimates, we obtain extra suppression of the scattering of the order
of $1/(D Y)$ for our case.

\subsection{Transverse momentum distribution in the BFKL Pomeron for the Gribov-Zwanziger confinement}

\begin{figure}
\begin{center}
\includegraphics[scale=1]{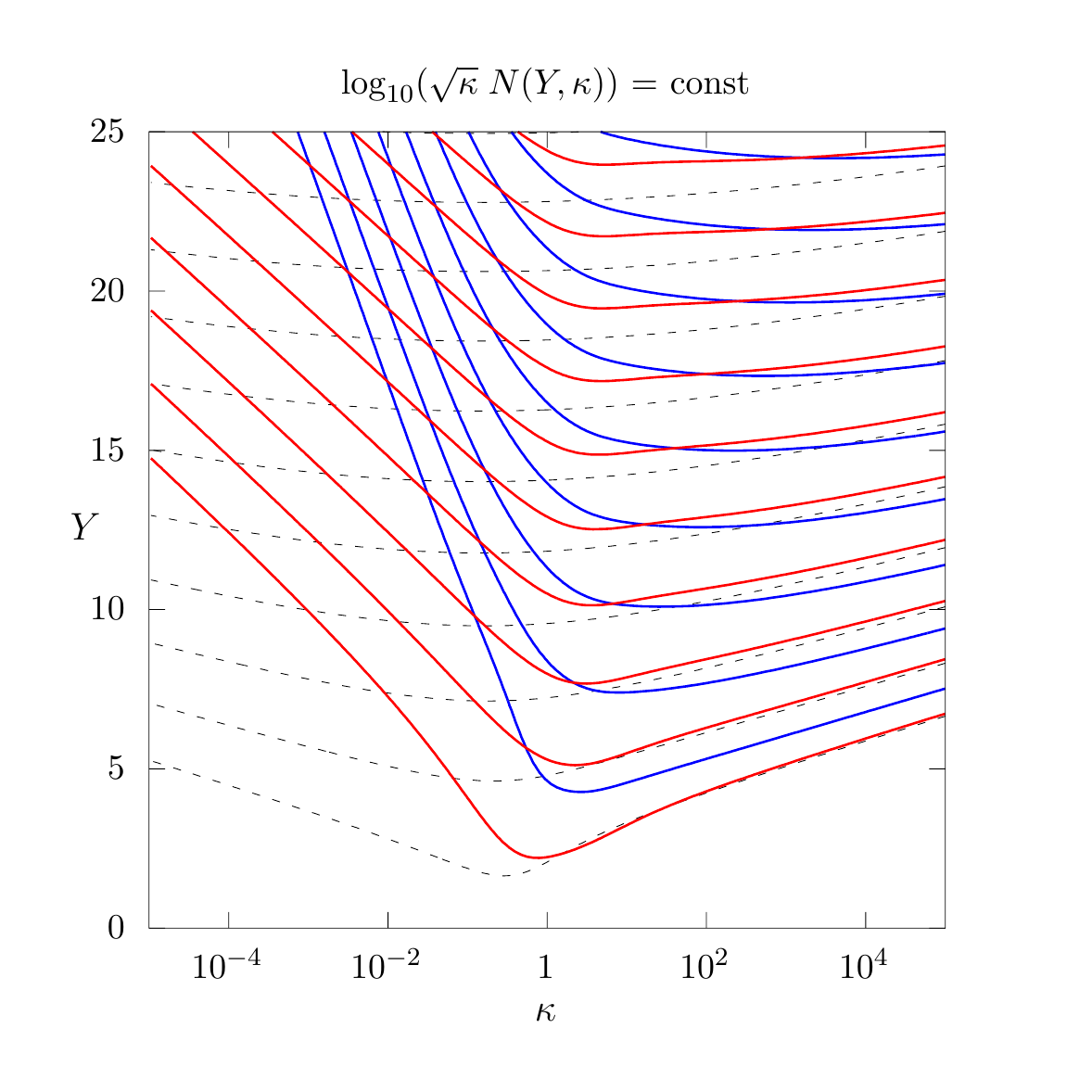}
\end{center}
\caption{\label{contour}
  The contour with constant $\kappa N\Lb Y, \kappa\Rb$ (see dashed line) for the QCD BFKL equation
  and for the BFKL equation for Gribov-Zwanziger confinement: the blue curve is for $m=0$ and the
  red one is for $m>0$.}
\end{figure}

Using \eq{GF1} we can find the scattering amplitude which will be equal to
\beq \label{TMD1}
  N\Lb Y; \kappa \Rb\,\,=\,\,\,\sum_{n=0}^{\infty}\,c_n\,e^{-\bas E_n\,Y} \, \phi_n(\kappa),
    \quad\mbox{with}\quad
    c_n \,=\, \int d \kappa_{\mbox{\tiny in}} \phi_n\Lb \kappa_{\mbox{\tiny in}}\Rb\, N\Lb Y=0,k_{\mbox{\tiny in}}\Rb
\eeq
where $N\Lb Y=0,\kappa_{\mbox{\tiny in}}\Rb$ is the initial condition for the scattering amplitude at
$Y=0$. In \fig{contour} $N\Lb Y=0,\kappa_{\mbox{\tiny in}}\Rb$ is taken to be equal to $1/(\kappa+1)^2$.

We plot in \fig{contour} the contours on which function $\kappa\,N\Lb Y,\kappa\Rb$
(see \eq{TMD1} is constant. In QCD we have transverse momentum distribution, which
depends on $|\ln\kappa\,|$,. The QCD evolution results in the increase of $|\ln\kappa\,|$
with $Y$. Such an increase leads to two possible branches (depending on different
sign of $\ln \kappa$) with increasing and decreasing average transverse momentum.
Such a behaviour is seen in \fig{contour}.

For our master equation one can see that the confinement cuts the small transverse momenta and
the average  $\kappa_T$ are larger than the values of $\kappa_T$ in initial conditions , which
we consider $\kappa_{\mbox{\tiny in}} = 1$, and they grow with $Y$. Therefore, introducing
the Gribov-Zwanziger confinement in the framework of the BFKL equation we obtain the transverse
momentum distribution, which is determined by the behaviour of the scattering amplitude at large
transverse momenta (at short distances), where we can trust the perturbative QCD approach.

\section{Conclusions}

In this paper we solved the new evolution equation for high energy scattering amplitude that stems
from the Gribov-Zwanziger approach to the confinement of quarks and gluons (see \eq{BFKLF1}). The
results of this solution we find quite surprising and instructive for future development of high
energy physics.

First, the energy dependence of the scattering amplitude turns out to be the same as for
QCD BFKL evolution. In particular, the eigenvalues of the new equation, which exceed
$\omega_0 = -\bas\,E_0 = -\bas\,T\Lb\kappa=0\Rb$, coincide with the QCD BFKL equation.
Second, the spectrum of the new equation does not depend on the details of the Gribov-Zwanzinger
approach and coincides with the set of the eigenvalues of the model: non-abelian gauge theories
with the Higgs mechanism for mass generation, developed in Ref.\cite{LLS}. This model has no
relation to a QCD approach except having the same colour structure. These features support
the ideas, that come out from the analytical analysis of the equation: the main influence
of the confinement is in taking off the double degeneration of the QCD BFKL equation, which
shows up in independence of the spectrum of the QCD BFKL equation on the sign of $\nu$
(see \eq{CHI}). Third, all eigenfunctions coincide with the eigenfunctions of the QCD BFKL
equation at large transverse momenta $\kappa\,\geq\,1$.

The numerical estimates show that there exist no new eigenvalues with the eigenfunctions that
decreases faster than the eigenfunction of the QCD BFKL equation at large transverse momenta.

The eigenfunctions of the master equation with the Gribov's gluon propagator tends to zero at
small transverse momenta. In the coordinate representation it means that the eigenfunctions
exhibit the power-like decrease at long distances, leading to the power-like decrease in the
impact parameters and, therefore, to the severe problem with Froissart theorem and s-channel
unitarity (see Refs.\cite{KW,FIIM,GOLEM}). In other words, the gluon propagator which tends
to zero as the Gribov's propagator does, cannot solve the problem with large $b$ dependence
of the scattering amplitude in the CGC approach. However, the structure of the gluon propagator
in Gribov-Zwanziger  approach that stems from the lattice QCD estimates and from the theoretical
evaluation (see Refs.\cite{HU1,CFPS,CFMPS,CDMV,DSV,HU2,HU3,AHS,DOV,GRA,FMP,DSVV,DGSVV,CLSST,Z4,Z5,LVS}),
leads to the gluon propagator which tends to a finite value at zero transverse momentum
($G\Lb q \to 0 \Rb\, \neq\, 0$). This results in the exponential suppression of the eigenfunction
at long distances and in the resolution of the difficulties, that the CGC approach as well as
other approaches, based on perturbative QCD, faces at large impact parameters.

For the intercept $\omega = -\bas\,T\Lb \kappa=0\Rb$ we have the multiple degeneration of this
eigenvalue, which is strongly correlated with the new dimensional parameter that we introduced
to the theory from the confinement. This degeneration looks as Bose-Einstein condensation but
it does not contribute to the scattering amplitude at high energy.

We calculate the momentum distributions of the scattering amplitude and found that the typical
transverse momentum increases with energy and become independent of the typical confinement scales
that we have introduced in our equation.

Therefore, to our surprise, we have to conclude that the confinement of quark and gluons,
at least in the form of Gribov-Zwanziger approach, does not influence on the scattering
amplitude except solving the long standing theoretical problem of the large impact parameter
behaviour of the scattering amplitude.

This is a very optimistic message for the CGC approach, but before coming to the strong conclusions
we have to check the solution to the non-linear equation with the new kernel. This will be our next
problem.

\section*{Acknowledgements}
We thank our colleagues at Tel Aviv University and UTFSM for encouraging discussions.
This research was supported by ANID PIA/APOYO AFB180002 (Chile) and FONDECYT (Chile)
grant 1180118.

\end{document}